\documentclass[10pt,notitlepage]{article}
\usepackage{graphicx}
\usepackage{amssymb}
= 0.0 in

\usepackage{amsmath}
\usepackage{amssymb}
\usepackage{amsthm}
\usepackage{graphicx}
\usepackage{mathrsfs}

\author{Achim Kempf\\\emph{ Department of Applied Mathematics, University of Waterloo}\\\emph{ Waterloo, Ontario, N2L 3G1, Canada}\\\\
Larissa Lorenz\\\emph{Institut d'Astrophysique de Paris, 98bis
boulevard Arago}\\\emph{ 75014 Paris, France}}
\title{Exact solution of inflationary model with minimum length}
\date{}   % Standard: Datum der Kompilierung ("\today").

\begin{document}

\maketitle
%\begin{center}
%\(^{\dagger}\)\emph{Department of Applied Mathematics, University of Waterloo,\\ Waterloo, Ontario, N2L 3G1, Canada}\\%{\bf akempf@uwaterloo.ca}\\
%\(^{\diamondsuit}\)\emph{Institut d'Astrophysique de Paris\\98bis
%boulevard Arago, 75014 Paris, France}%\\{\bf lorenz@iap.fr}
%\end{center}
\begin{abstract}
Within the inflationary scenario, Planck scale physics should have
affected the comoving modes' initial conditions and early evolution,
thereby potentially affecting the inflationary predictions for the cosmic
microwave background (CMB). This issue has been studied extensively on the
basis of various models for how quantum field theory (QFT) is modified and
finally breaks down towards the Planck scale. In one model, in particular,
an ultraviolet cutoff was implemented into QFT through generalized
uncertainty relations which have been motivated from general quantum
gravity arguments and from string theory. Here, we improve upon prior
numerical and semi-analytical results by presenting the exact mode
solutions for both de Sitter and power-law inflation in this model. This
provides an explicit map from the modes' initial conditions, which are
presumably set by quantum gravity, to the modes' amplitudes at horizon
crossing and thus to the inflationary predictions for the CMB. The
solutions' particular behaviour close to the cutoff scale suggests
unexpected possibilities for how the degrees of freedom of QFT emerge from
the Planck scale.

\end{abstract}

%\tableofcontents

\twocolumn

\section{Introduction}\label{introduction}
Within the inflationary scenario, spacetime typically inflated to the
extent that comoving modes which are of cosmological size today originated
well beyond the Planck scale (and any other natural ultraviolet (UV)
cutoff scale such as the string scale). This observation has led to a
growing body of work exploring the tantalizing prospect that the
predictions of inflation could be observably affected by quantum gravity.
For early papers on this subject see, e.g.,
\cite{brandenbergermartin01-a,brandenbergermartin01-b,niemeyer01,kempf00,kempfniemeyer01}.

Technically, all possible Planck scale effects on the comoving modes, $k$,
 at late times take the form of a nontrivial $k$-dependent selection of
solution to the ordinary mode equation, see, e.g., \cite{starobinsky01}.
This is because, independent of the details of Planck scale physics, each
mode's evolution equation reduces to the ordinary wave equation as soon as
its proper wave length is significantly larger than the Planck scale. It
has been possible, therefore, to determine on these general grounds that
quantum gravity effects should manifest themselves qualitatively in the
form of superimposed oscillations in the CMB spectra as well as in a
breaking of the consistency relation between the scalar$/$tensor ratio and
the tensor spectral index of single field inflation, see, e.g.,
\cite{greeneetal0104,huikinney01,greeneetal0110,greeneetal02,jerome-osc}.

Any quantitative predictions, however, are particular to each particular
theory of quantum gravity. Concretely, for wavelengths closer and closer
to the cutoff scale the framework of QFT should hold - with characteristic
increasing corrections - until it finally breaks down at the cutoff scale.
A theory of quantum gravity, be it, e.g., loop quantum gravity or $M$
theory, should eventually allow one to calculate those characteristic
corrections to the framework of QFT in this ``sub-Planckian" regime, i.e.,
in the regime of wavelengths larger than but close to the cutoff length.
In addition, the theory of quantum gravity should allow one to determine
the initial condition for the evolution of a comoving mode within the
framework of QFT, namely at the time when the mode's proper wavelength
first exceeds the cutoff length.

In this context, first-principle calculations are very difficult, of
course. Therefore, a number of authors have modelled the corrections to
the framework of QFT from quantum gravity simply as UV modifications to
the dispersion relations. The implied effects on the inflationary
predictions for the CMB have been investigated, for example, in
\cite{brandenbergermartin01-a,brandenbergermartin01-b,niemeyer01,niemeyerparentani01}.
Particular modifications to the dispersion relations for short wavelengths
have been motivated either \emph{ad hoc} or on the basis of analogies to
the propagation of waves in condensed matter systems. It was found that
the effects of Planck scale physics could be considerable, even though
there appear to be strong constraints due to a possibly strong
backreaction problem, see, e.g., \cite{tanaka00}.

A less \emph{ad hoc} approach \cite{kempf00} has been to model the
behaviour of quantum field theory in the sub-Planckian regime through the
implementation of corrections to the first quantization uncertainty
relations, see \cite{kempf98}. These corrections implement an ultraviolet
cutoff in the form of a finite minimum uncertainty in spatial distances,
which has been motivated by studies in quantum gravity and string theory,
see, e.g.,
\cite{Gross:1987ar,Amati:1988tn,Garay:1994en,Amelino-Camelia:1997uq,Witten:2001ib}.

This approach to modelling Planck scale physics in inflation was further
investigated in particular in
\cite{kempfniemeyer01,greeneetal0104,greeneetal0110,greeneetal02}. So far
it has been possible to solve only approximations of the mode equation
that arises in this approach. In fact, to solve the exact mode equation is
known to be highly nontrivial even numerically because of a particular
singular behaviour of the mode equation at the mode's starting time. Here,
we are following up on this series of papers.

In particular, we calculate here the exact solutions to the exact mode
equation in both the de Sitter and power-law backgrounds. This allows us
then to follow the mode solutions back towards the time when the proper
wavelength approaches the cutoff length. That in turn allows us to
calculate the exact behaviour of physical quantities such as the field
fluctuation amplitudes and the Hamiltonian towards the Planckian regime.
We obtain, therefore, a fully explicit model in which to explore possible
mechanisms that could impose initial conditions on comoving modes as they
enter into the regime of validity of quantum field theory.

\section{Short distance physics and inflation}\label{shortdistance}
\subsection{UV cutoff through minimum length uncertainty}
A rather general assumption about quantum field theory in the
sub-Planckian regime is that it should still possess for each coordinate a
linear operator \({\bf x}^{i}\) (inherited from first quantization) whose
formal expectation values (e.g., in the space of fields that are being
summed over in the path integral) are real. The \({\bf x}^{i}\) may or
 may not commute. As shown in \cite{kempf98}, this implies that the short distance
  structure of any such coordinate, considered separately,
 can only be continuous, discrete, or ``unsharp'' in one of two particular ways.
Studies in quantum gravity and string theory point towards one of these
two unsharp cases, namely the case of coordinates \({\bf x}\) whose formal
uncertainty \(\Delta {\bf x}$ possesses a finite lower bound $\Delta {\bf
x}_{min}$ at the Planck or string scale. This short distance structure
arises from quantum gravity correction terms to the uncertainty relation
(see \cite{uncertainty}):
\begin{equation}%\label{corrcommutation}
\Delta x\Delta p\geq \frac{1}{2}(1+\beta(\Delta p)^{2}+\dots)
\end{equation}
Here, \(\beta\) is a positive constant which, as is easily verified,
parametrizes the cutoff length through \(\Delta x_{min}=\sqrt{\beta}\).
The units are such that $\hbar=1$. As was first pointed out in
\cite{kempf93}, uncertainty relations of this form arise from corrections
to the canonical commutation relations which, for example in the case of
one space dimension, take this form:
\begin{equation}\label{corrcommutation}
 [{\bf x},{\bf p}]=i(1+\beta {\bf p}^{2}+\dots)
\end{equation}
It was shown in \cite{ak-harmosc} that the multi-dimensional
generalization is unique to first order in $\beta$ if rotational and
translational symmetry is to be preserved.

Let us remark that, as was shown in
 \cite{ak-sampling},  this type of natural UV cutoff implies
that physical fields possess a finite bandwidth in the information
theoretic sense: In the presence of this short distance structure one can
apply Shannon's sampling theorem to conclude that if the values of a
physical field are known on any set of points with average density above
twice the Planck density then the values of the field everywhere are
already determined and can be explicitly calculated from these
samples\footnote{The situation is mathematically identical, for example,
to that of HiFi music signals of bandwidth 20KHz, whose amplitude samples
are recorded on CDs at a rate of $4\cdot 10^4$ samples per second. Using
Shannon's theorem, CD players are able to \it precisely \rm reconstruct
the continuous music signal at \it all \rm time from these discrete
samples.}. Notice that no sampling lattice is preferred, as all sampling
lattices of sufficiently high  average density can be used to reconstruct
the fields. Therefore, in a theory with this type of natural cutoff
spacetime can be viewed as continuous, in which case the conservation of
the spatial symmetries is displayed, while, fully equivalently, in the
same theory spacetime can also be viewed as discrete, in which case the
theory's UV finiteness is displayed. So far, this type of natural UV
cutoff has been considered mostly within models that break Lorentz
invariance, as we will here. The generalization to a generally covariant
setting has been started in \cite{ak-sampling-cov}.

 %\subsection{Inflation as a Planck scale microscope}

%Before introducing the formalism used here, let us mention that recent work \cite{amjad} has shown that Planck scale physics might be influential in cosmological perturbation theory through a more subtle mechanism as well: In the presence of a cutoff, the action for cosmological perturbations is no longer insensitive to the addition or subtraction of terms which would be mere total time derivatives for the cutoff length set to zero. (This can be understood analogously to the usual ordering ambiguity of operators in quantum mechanics.) Therefore two formerly equivalent formulations of the equation of motion are rendered different by the UV cutoff.

\subsection{Mode generation in the presence of the UV cutoff}
In \cite{kempf00}, the short distance cutoff formulated in
(\ref{corrcommutation}) was applied to the theory of a minimally coupled
massless real scalar field \(\phi({\bf x},t)\) in an expanding
Friedmann-Robertson-Walker (FRW) background. This setup can be used to
describe the quantum dynamics of the tensor as well as the scalar
fluctuations in inflation. Concerning the mode evolution, the only
difference is that, in the latter case, the scale factor $a$ in the mode
equation is replaced by $z=\phi_0'a^2/a'$. Here, $\phi_0$ is the bulk zero
mode of the inflaton field with the prime denoting differentiation with
respect to \(\eta\). (It should be kept in mind, however, that the
Hamiltonian of the gravitational waves that are due to tensor modes
contributes to the energy momentum tensor only from second order.)

The approach of \cite{kempf00} was continued in \cite{kempfniemeyer01} as
well as in a series of papers by Easther \emph{et al.}
\cite{greeneetal0104,greeneetal0110,greeneetal02}. In each case, for
simplicity, the notation for the tensor modes was used (i.e. with the mode
equation containing $a$ instead of $z$), as we will also do here.

We assume the case of spatial flatness of the background spacetime and we
use the conformal time coordinate \(\eta\). Note that the UV cutoff
\(\Delta x_{min}=\sqrt{\beta}\)
 is introduced in proper distances as opposed to comoving distances.
The action for the field \(\phi({x}, \eta)\) then reads, see
\cite{kempf00}:
\begin{eqnarray}
S&=&\int d\eta d^{3}x\,\frac{1}{2a}\bigg\{\left[\left(\partial_{\eta}+\frac{a'}{a}\sum_{i=1}^{3}\partial_{x^{i}}x^{i}-\frac{3a'}{a}\right)\phi\right]^{2}\nonumber\\
&&-a^{2}\sum_{i=1}^{3}\left(\partial_{x^{i}}\phi\right)^{2}\bigg\}\label{action}
\end{eqnarray}
Note that our function \(\phi_{\tilde{k}}(\eta)\) is related to
\(u_{\tilde{k}}\) in \cite{greeneetal0104} by
%\footnote{Note that due to a
%typo, the relation given in \cite{greeneetal0104} reads
%\(u_{\tilde{k}}=a^{2}\cdot \phi_{\tilde{k}}(\eta)\).}
\(u_{\tilde{k}}=1/a^{2}\cdot \phi_{\tilde{k}}(\eta)\) and that our
function \(\nu(\eta,\tilde{k})\) is the function \(\nu(\eta,\rho)\) from
\cite{greeneetal0104} multiplied by \(a^{4}\).
Using the operators $${\bf x}^{i}: \phi(x,\eta)\rightarrow x_i\phi(x,\eta)$$
$${\bf p}^{i}: \phi(x,\eta)\rightarrow -i\partial/\partial x_i\phi(x,\eta)  $$ %(time dependent abstract vectors in a Hilbert space representation of the usual commutation relations \(\left[{\bf x}^{i},{\bf p}^{j}\right]=i\delta^{ij}\))
the action can be expressed representation independently in terms of these
operators and the usual $L^2$ inner product of fields. While the fields'
expression for the action (\ref{action}) is held fixed the underlying
commutation relations are now corrected to introduce the desired cutoff
(\ref{corrcommutation}). The modification of the short distance behaviour
then manifests itself whenever one writes the action in any
representation, such as in comoving momentum space. For example, at
distances close to the cutoff length, comoving modes no longer decouple.
However, it is then still possible to find variables \((\eta,\tilde{k})\)
in which the \(\tilde{k}\) modes of the fields' Fourier transform do
decouple\footnote{Note that, as discussed in \cite{kempf00} and
\cite{kempfniemeyer01}, these variables \(\tilde{k}\) approximate the
comoving momentum variables \(k\) only at long wavelengths.}. The
decoupling modes are important as they describe the truly independent
degrees of freedom. They will be denoted as \(\phi_{\tilde{k}}(\eta)\).
The realness of the field \(\phi({\bf x},\eta)\) translates into
\(\phi_{\tilde{k}}^{*}=\phi_{-\tilde{k}}\). The action (\ref{action}) can
then be written in the form
\begin{equation}S=\int
d\eta\int_{\tilde{k}^{2}<a^{2}/e\beta}d^{3}\tilde{k}\,\mathscr{L}\label{af}\end{equation}
with the Lagrangian
\begin{equation}\label{lagrangian}
\mathscr{L}=\frac{1}{2}\nu\left\{\left|\left(\partial_{\eta}-3\frac{a'}{a}\right)\phi_{\tilde{k}}(\eta)\right|^{2}-\mu\left|\phi_{\tilde{k}}(\eta)\right|^{2}\right\},
\end{equation}
where the coefficient functions are given by
\begin{eqnarray}
\mu(\eta,\tilde{k})&:=&-\frac{a^{2}\textnormal{plog}(-\beta\tilde{k}^{2}/a^{2})}{\beta\left[1+\textnormal{plog}(-\beta\tilde{k}^{2}/a^{2})\right]^{2}},\label{mu}\\
\nu(\eta,\tilde{k})&:=&\frac{\exp\left[-\frac{3}{2}\textnormal{plog}(-\beta\tilde{k}^{2}/a^{2})\right]}{a^{4}\left[1+\textnormal{plog}(-\beta\tilde{k}^{2}/a^{2})\right]}.\label{nu}
\end{eqnarray}
The function plog used here, the ``product log'', is defined as the
inverse of the function \(x\rightarrow xe^{x}\) (also known as the Lambert
W-function).  It will be important later that, at \(x=-1/e\), this
function possesses an essential singularity.

Equally important will be the way in which the short distance
cutoff\footnote{For \(\beta\rightarrow 0\), (\ref{lagrangian}) reduces to
the standard action. Note that Fourier transforming and scaling only
commute up to a scaling factor of \(a^{3}\).} affects the integration
region of the action functional (\ref{af}): each mode \(\tilde{k}\) enters
the action once \(a(\eta)\) has grown enough so that the condition
\(\tilde{k}^{2}<a^{2}/e\beta\) holds, i.e., when:
\begin{equation*}
a(\eta_{c})=\tilde{k}\sqrt{e\beta}\approx\tilde{k}\Delta x_{min}.
\end{equation*} Correspondingly,
the mode equation that follows from the action (\ref{lagrangian}), see
\cite{kempf00}, {\small
\begin{equation}\label{equationofmotion}
\phi_{\tilde{k}}''+\frac{\nu'}{\nu}\phi_{\tilde{k}}'+\left[\mu-3\left(\frac{a'}{a}\right)'-9\left(\frac{a'}{a}\right)^{2}-3\frac{a'\nu'}{a\nu}\right]\phi_{\tilde{k}}=0
\end{equation}}
is an equation of motion which possesses a starting time. As had to be
expected, the implementation of a UV cutoff in an expanding spacetime has
led us to the problem  of setting initial conditions on these newly
emerging comoving modes. Before we solve the mode equation to obtain the
solutions' early time behavior, let us briefly consider the mode
solutions' late time behaviour.

\subsection{Late time behaviour}

At late times, when a mode's proper wavelength far exceeds the cutoff
length, the mode becomes insensitive to the nontrivial short distance
structure of spacetime. The modified mode equation
(\ref{equationofmotion}) then reduces to the mode equation without cutoff,
which (taking into account the non-com\-mutativity of scaling and Fourier
transforming)
reads with our conventions\footnote{Recall that at large scales we also have \(\tilde{k}\rightarrow k\).}:%That is to say, when the wavelength is big enough, the mode should not care about the discrete structure of space time. Equation (\ref{equationofmotion}) is different from the usual formulation of the equation of motion not only because of the cutoff, but also because the field  \(\phi_{\tilde{k}}(\eta)\) differs from the usual field  \(\Phi_{\tilde{k}}(\eta)\) by a factor of \(a^{-3}(\eta)\) (non-commutativity of scaling and Fourier transforming, see \cite{kempf00}). Therefore the usual mode equation for \(\phi_{\tilde{k}}(\eta)\) without a cutoff reads
\begin{equation}\label{philatetime}
\phi_{\tilde{k}}''-4\frac{a'}{a}\phi_{\tilde{k}}'+\left(6\left(\frac{a'}{a}\right)^{2}-3\frac{a''}{a}+\tilde{k}^{2}\right)\phi_{\tilde{k}}=0
\end{equation}
%There will be some point in time \(\eta_{0}\) from when on the difference between the modified and this usual equation will no longer be significant. We will first examine the behaviour of (\ref{philatetime}) for \(\eta\rightarrow0\) and then investigate the criteria under which (\ref{equationofmotion}) reduces to (\ref{philatetime}).
%\subsection{Assymptotic behaviour for late times}
Let us derive the late time behaviour of (\ref{philatetime}), namely
\(\eta\rightarrow0\), in de Sitter space \(a(\eta)=-1/H\eta\). Since in
this case we have \(\frac{a'}{a}=-\frac{1}{\eta}\) and
\(\frac{a''}{a}=\frac{2}{\eta^{2}}\), while \(\tilde{k}^{2}\) is constant,
equation (\ref{philatetime}) on large scales becomes:
\begin{eqnarray*}
%\phi_{\tilde{k}}''-4\frac{a'}{a}\phi_{\tilde{k}}'+\left(6\left(\frac{a'}{a}\right)^{2}-3\frac{a''}{a}\right)\phi_{\tilde{k}}&=&\nonumber\\
\phi_{\tilde{k}}''+4\frac{1}{\eta}\phi_{\tilde{k}}'&=&0
\end{eqnarray*}
The solution to this equation is given by
\begin{equation}
\phi_{\tilde{k}}=B_{1}+\frac{B_{2}}{\eta^{3}}.
\end{equation}
So for late times, \(\eta\rightarrow0\), the mode function
\(\phi_{\tilde{k}}(\eta)\) diverges \(\propto\frac{1}{\eta^{3}}\) in the
de Sitter case. We recall that
\(\phi_{\tilde{k}}(\eta)=a^{3}(\eta)\Phi_{\tilde{k}}(\eta)\). Thus, as
expected, the physical field's mode function \(\Phi_{\tilde{k}}(\eta)\)
approaches a constant for late times. Let us remark that the analogous
calculation for classical fluctuations indicates that classical scalar or
tensor fluctuations that existed before inflation would have their
wavelength stretched but their amplitude maintained, i.e., they would not
be ``ironed out" by inflation. Inflation is usually thought to have lasted
long enough so that, classically, the inflation-induced flattening of a
fixed proper spatial volume would hinge on the assumption that there were
no pre-existing spatial ``ripples" down to wavelengths shorter than the
Planck length. Of course, the Planck scale is expected to rule out the
notion of pre-existing classical ripples that are shorter than the Planck
length.

\section{The Hamiltonian}\label{Hamiltoniananalysis}
Using the canonically conjugate field to $\phi_{\tilde{k}}$, see
\cite{kempf00},
\(\pi_{\tilde{k}}(\eta)=\nu\phi'_{-\tilde{k}}(\eta)-3\nu\frac{a'}{a}\phi_{-\tilde{k}}(\eta)\),
the Hamiltonian
%\(H=\int_{\tilde{k}^{2}<a^{2}/e\beta}d^{3}\tilde{k}\left[\pi_{\tilde{k}}(\eta)\phi'_{\tilde{k}}(\eta)-\mathscr{L}\right]\)
corresponding to the action (\ref{action}) takes the form
\begin{equation}\label{Hamiltonian}
H=\int_{\tilde{k}^{2}<\frac{a^{2}}{e\beta}} d^{3}\tilde{k}
\left(\frac{1}{2\nu}\pi_{\tilde{k}}^{*}\pi_{\tilde{k}}+\frac{1}{2}
\nu\mu\,\phi_{\tilde{k}}^{*}\phi_{\tilde{k}}+3\frac{a'}{a}\pi_{\tilde{k}}
\phi_{\tilde{k}}\right).
\end{equation}
To quantize one promotes \(\hat{\phi}_{\tilde{k}}\) and
\(\hat{\pi}_{\tilde{k}}\) to operators satisfying the usual commutation
relation
\([\hat{\phi}_{\tilde{k}},\hat{\pi}_{\tilde{r}}]=i\delta^{3}(\tilde{k}-\tilde{r})\)
in momentum space. Note that the field commutators are of course modified
when expressed in position space.
%(The Heisenberg equations \(\hat{\phi}'_{\tilde{k}}=i\left[\hat{H},\hat{\phi}_{\tilde{k}}\right],\hat{\pi}'_{\tilde{k}}=i\left[\hat{H},\hat{\pi}_{\tilde{k}}\right]\) then give (\ref{equationofmotion}) as an equation for operator valued fields.)

In the Hamiltonian, the integration region in\-crea\-ses over time,
expressing that the field operators of comoving modes enter the
Hamiltonian only when the mode's wavelength exceeds the cutoff length.
This is significant because it means that the Hamiltonian contains
operators which act nontrivially on the Hil\-bert space dimensions of a
particular comoving mode \(\tilde{k}\) only from that time
\(\eta_{c}(\tilde{k})\) when this mode's proper wavelength starts to
exceed the cutoff length. In the literature, this time (which coincides
with the essential singularity of the plog function) has been called the
mode's ``creation time'' and we will stay with this terminology. Let us
keep in mind, however, that the dimensions on which the harmonic
oscillator of a given mode is represented are of course \it always \rm in
the Hilbert space. In the Heisenberg picture, all that happens at the
creation time of a mode is that the mode's field and conjugate field
operators enter the Hamiltonian for the first time, while conversely, in a
shrinking universe, a mode's field operators would drop from the
Hamiltonian when the mode's wavelength drops below the cutoff length.

We express the quantum field in terms of creation and annihilation
operators \(a_{\tilde{k}}^{\dagger}\) and \(a_{\tilde{k}}\) and in terms
of the mode functions \(\phi_{\tilde{k}}(\eta)\):
\begin{equation}
\hat{\phi}_{\tilde{k}}(\eta)=a_{\tilde{k}}\phi_{\tilde{k}}(\eta)+a_{-\tilde{k}}^{\dagger}\phi_{-\tilde{k}}^{*}(\eta),
\end{equation}
Note that the notation \(\phi_{\tilde{k}}(\eta)\) now no longer stands for
the classical field (which obeys $\phi^*_{\tilde{k}}=\phi_{-k}$) but
instead for a mode function. For the mode functions we can choose, as
usual that \(\phi_{\tilde{k}}=\phi_{\tilde{|k|}}\).
 % with \([a_{\tilde{k}}, a_{\tilde{r}}^{\dagger}]=\delta^{3}(\tilde{k}-\tilde{r})\) and use them to express the operator \(\hat{\phi}_{\tilde{k}}\) through the classical field \(\phi_{\tilde{k}}\): \(\hat{\phi}_{\tilde{k}}(\eta)=a_{\tilde{k}}\phi_{\tilde{k}}(\eta)+a_{-\tilde{k}}^{\dagger}
%\phi_{-\tilde{k}}^{*}(\eta)\) Note that then in order for the second quantization commutation relations for \(\hat{\phi}_{\tilde{k}}\) and \(\hat{\pi}_{\tilde{k}}\) to hold, we have to impose
The Wronskian condition%\(\phi_{\tilde{k}}\),
\begin{equation}\label{wronskian}
\nu(\eta,\tilde{k})\left[\phi_{\tilde{k}}(\eta)\phi'^{*}_{\tilde{k}}(\eta)-\phi^{*}_{\tilde{k}}(\eta)\phi'_{\tilde{k}}(\eta)\right]=i.
\end{equation}
insures that the so-defined quantum fields obey the canonical commutation
relations.
%on the classical fields which solve (\ref{equationofmotion}). (Note that the  \(a_{\tilde{k}}^{\dagger}\) and
%\(a_{\tilde{k}}\) are time independent.) The realness of the classical field also gives \(\left|\phi_{\tilde{k}}\right|^{2}=\left|\phi_{-\tilde{k}}\right|^{2}\).
%We can calculate the operator \(\hat{\pi}_{\tilde{k}}\) in terms of creation and annihilation operators: \(\hat{\pi}_{\tilde{k}}=\nu\hat{\phi}_{-\tilde{k}}'-3\nu\frac{a'}{a}\hat{\phi}_{-\tilde{k}}
%=\left(\nu\phi_{-\tilde{k}}'-3\nu\frac{a'}{a}\phi_{-\tilde{k}}\right)a_{-\tilde{k}}
%+\left(\nu\phi_{\tilde{k}}^{'*}-3\nu\frac{a'}{a}\phi_{\tilde{k}}^{*}\right)a_{\tilde{k}}^{\dagger}\).
%If we insert \(\hat{\phi}_{\tilde{k}}\) and \(\hat{\pi}_{\tilde{k}}\) in this form into the quantized Hamiltonian that follows from (\ref{Hamiltonian}), combined terms of corresponding \(\tilde{k}\)- and \(-\tilde{k}\)-modes and write terms as Hermitian or complex conjugates of each other where possible, it simplifies to
%Writing  \(\hat{\phi}_{\tilde{k}}\) and \(\hat{\pi}_{\tilde{k}}\) in terms of \(a_{\tilde{k}}^{\dagger}\) and
%\(a_{\tilde{k}}\) and the classical fields,
We can then rearrange the quantized Hamiltonian in the following form that
we will later refer to: {\small
\begin{eqnarray}
\hat{H}&=&\int_{\tilde{k}^{2}<\frac{a^{2}}{e\beta}} d^{3}\tilde{k}
\left[\frac{\nu}{2}\left|\phi_{\tilde{k}}'\right|^{2}+
\frac{\nu}{2}\left(\mu-9\left(\frac{a'}{a}\right)^{2}\right)
\left|\phi_{\tilde{k}}\right|^{2}
\right]\nonumber\\
&&\times\left(a_{\tilde{k}}^{\dagger}a_{\tilde{k}}+
a_{-\tilde{k}}^{\dagger}a_{-\tilde{k}}\right)
\nonumber\\
&+&\left[\frac{\nu}{2}\phi_{-\tilde{k}}'\phi_{\tilde{k}}'+
\frac{\nu}{2}\left(\mu-9\left(\frac{a'}{a}\right)^{2}\right)
\phi_{-\tilde{k}}\phi_{\tilde{k}}\right]\nonumber\\
&&\times \left(a_{\tilde{k}}a_{-\tilde{k}}\right)\nonumber\\
&+&\left[\frac{\nu}{2}\left(\phi_{-\tilde{k}}'\phi_{\tilde{k}}'\right)^{*}+
\frac{\nu}{2}\left(\mu-9\left(\frac{a'}{a}\right)^{2}\right)
\left(\phi_{-\tilde{k}}\phi_{\tilde{k}}\right)^{*}\right]\nonumber\\
&&\times\left(a_{\tilde{k}}a_{-\tilde{k}}\right)^{\dagger}\nonumber\\
&+&\left[\frac{\nu}{2}\left|\phi_{\tilde{k}}'\right|^{2}+
\frac{\nu}{2}\left(\mu-9\left(\frac{a'}{a}\right)^{2}\right)
\left|\phi_{\tilde{k}}\right|^{2}\right]\delta^{3}(0).\label{quantizedHamiltonian}
\end{eqnarray}}

To see that this Hamiltonian gives the correct equation of motion (\ref{equationofmotion}), it is straightforward to apply the Heisenberg equations and use the definition of \(\hat{\pi}_{\tilde{k}}\). Note again the bound on the integral, which reflects the fact that each mode \(\tilde{k}\) only contributes to the Hamiltonian from its creation time \(\eta_{c}(\tilde{k})\) onwards; before that time, the operators \(a_{\tilde{k}}\) and \(a^{\dagger}_{\tilde{k}}\) are not contained in \(\hat{H}\).\\

Let us note that, curiously, in the case of a shrinking universe the
modes' creation and annihilation operators \(a_{\tilde{k}},
a^{\dagger}_{\tilde{k}}\) successively drop out of the Hamiltonian. Thus,
a growing number of states can no longer be ``reached'' by the Hamiltonian
and the field operators while the universe is contracting. All information
encoded in these states would no longer interact although it would become
relevant again during a subsequent expanding phase.

We will be interested, in particular, in the ground state energy term of
each comoving mode, given by the last line of
(\ref{quantizedHamiltonian}):
\begin{eqnarray}
\label{vacuumterm}
\rho_{vac,\tilde{k}}(\eta)&=&\frac{\nu(\eta)}{2}
\left|\phi_{\tilde{k}}'(\eta)\right|^{2}\\
&+&
\frac{\nu(\eta)}{2}\left|\phi_{\tilde{k}}(\eta)\right|^{2}\left(\mu(\eta)-9
\left(\frac{a'(\eta)}{a(\eta)}\right)^{2}\right)
\nonumber
\end{eqnarray}
Since the Hamiltonian is closely related to the \(00\) component of the
energy momentum tensor \(T^{\mu}{}_{\nu}\), (\ref{vacuumterm}) essentially
describes the energy contribution of each mode which arises when the mode
outgrows the cutoff scale. Two time dependencies determine the evolution
of the modes' ground state energy (\ref{vacuumterm}): first, the behaviour
of the coefficients \(\nu(\eta,\tilde{k})\) and \(\mu(\eta,\tilde{k})\),
and second the mode function \(\phi_{\tilde{k}}(\eta)\). While the former
are already known explicitly (see (\ref{mu}) and (\ref{nu})), as we will
discuss, the choice of solution to the equation of motion
(\ref{equationofmotion}) for the mode function will nontrivially affect
the ground state energy.

\section{Solving the mode equation}\label{eofmanalysis}
We will now derive the exact solutions to (\ref{equationofmotion}) which
obey the Wronskian condition (\ref{wronskian}). The difficulty here is due
to the plog-functions in \(\mu(\eta,\tilde{k})\) and
\(\nu(\eta,\tilde{k})\): the equation of motion for each
\(\tilde{k}\)-mode has an irregular singular point at the mode's creation
time \(\eta_{c}(\tilde{k})\). In previous work, \cite{greeneetal0104}, the
mode equation was replaced by an approximate mode equation which had a
regular singular point at the mode's creation time. The exact solutions to
this approximate mode equation were obtained both as power series and in
analytical form.

Our aim now is to obtain the exact solutions to the exact mode equation.
To this end, we employ a suitable variable transformation that turns the
irregular singular point into a regular singular point, without the need
for approximations. We then apply the Frobenius method to derive the exact
solutions, namely in terms of convergent power series and logarithms.

\subsection{Singularity transformation}
%In the de Sitter case \(a(\eta)=-\frac{1}{H\eta}\), each mode is created at the time
%\begin{equation}\label{creationdesitter}
%\eta_{c}(\tilde{k})=-\frac{1}{H\tilde{k}\sqrt{e\beta}}
%\end{equation}
%when its wavelength first equals the minimum length.
Technically, \(\eta_{c}(\tilde{k})\) is called an irregular singular point
of (\ref{equationofmotion}) because the term
\begin{equation*}
\left(\eta-\eta_{c}(\tilde{k})\right)^{2}
\left[\mu-3\left(\frac{a'}{a}\right)'-9\left(\frac{a'}{a}\right)^{2}-3\frac{a'\nu'}{a\nu}\right]
\end{equation*}
and the term
\begin{equation*}
\left(\eta-\eta_{c}(\tilde{k})\right)\cdot\frac{\nu'}{\nu}
\end{equation*}
do not have convergent Taylor expansions around this point. Let us now
define a new variable \(\tau\),
\begin{equation}\label{deftau}
\tau=\textnormal{plog}(-\beta\tilde{k}^{2}/a(\eta)^{2}),
\end{equation}
which plays the role of time, but is a function of both \(\eta\) and
\(\tilde{k}\).
%In the de Sitter case, this leads to the inverse transformation
%\begin{equation}\label{etaoftau}
%\eta=-\frac{1}{H\tilde{k}}\frac{\sqrt{-\tau e^{\tau}}}{\sqrt{\beta}}.
%\end{equation}
%After the transformation \(\eta\rightarrow\tau\)
The scale factor \(a(\eta)\) and the coefficient functions
\(\nu(\eta,\tilde{k})\) and \(\mu(\eta,\tilde{k})\) then
read\footnote{Note that to distinguish between functions in \(\eta\) as
opposed to functions in \(\tau\), we will write a tilde on top of
functions in \(\tau\), so \(a=a(\eta)\), but \(\tilde{a}=\tilde{a}(\tau)\)
etc. and especially \(\phi_{\tilde{k}}=\phi_{\tilde{k}}(\eta)\) but
\(\tilde{\phi}_{\tilde{k}}=\tilde{\phi}_{\tilde{k}}(\tau)\). This should
not be confused with the tilde on \(\tilde{k}\) which was introduced in
\cite{kempf00} to denote coordinates in which the Fourier modes decouple.}
\begin{eqnarray}
\tilde{a}(\tau)&:=&\tilde{k}\sqrt{\frac{\beta}{-\tau e^{\tau}}},\label{aoftau}\\
\tilde{\mu}(\tau,\tilde{k})&:=&\frac{\tilde{k}^{2}}{e^{\tau}(1+\tau)^{2}},\label{muoftau}\\
\tilde{\nu}(\tau,\tilde{k})&:=&\frac{\tau^{2}e^{\frac{\tau}{2}}}{\beta^{2}\tilde{k}^{4}(1+\tau)}.\label{nuoftau}
\end{eqnarray}
It is straightforward to translate the derivatives $\partial /\partial \eta$ into derivatives in
terms of derivatives \(\partial /\partial\tau\) and we then obtain that the equation of motion (\ref{equationofmotion}) takes the form:
\begin{eqnarray}
0  & = & \tilde{\phi}_{\tilde{k}}''+\left(\frac{\tilde{\nu}'}{\tilde{\nu}}-\frac{\eta''}{\eta'}\right)
\tilde{\phi}_{\tilde{k}}' \label{transformedeofm}\\
&  & + \bigg[\tilde{\mu}\,\eta'^{2}-3\left(\frac{\tilde{a}''}{\tilde{a}}\right)- 6\left(\frac{\tilde{a}'}{\tilde{a}}\right)^{2} \nonumber \\
& & ~~~-
3\frac{\tilde{a}'\tilde{\nu}'}{\tilde{a}\tilde{\nu}}+3\frac{
\tilde{a}'}{\tilde{a}}\frac{\eta''}{\eta'}\bigg]\tilde{\phi}_{\tilde{k}}\nonumber
\end{eqnarray}
Note that we now let the prime denote differentiation with respect to \(\tau\) rather than \(\eta\).

In order to make those expressions explicit in an example, let us consider the case of de Sitter space
\(a(\eta)=-1/H\eta\). In this case, by inverting (\ref{deftau}), $\eta$ is expressed in terms of $\tau$ and $\tilde{k}$ through:
\begin{equation}\label{etaoftau}
\eta(\tau,\tilde{k})=-\frac{1}{H\tilde{k}}\frac{\sqrt{-\tau
e^{\tau}}}{\sqrt{\beta}}
\end{equation}
Using (\ref{aoftau}), (\ref{muoftau}) and (\ref{nuoftau}), we then obtain for the mode equation in de Sitter space:
\begin{equation}\label{eofmexplicit}
\tilde{\phi}_{\tilde{k}}''+\frac{5+\tau}{2\tau(1+\tau)}
\tilde{\phi}_{\tilde{k}}'
-\left(\frac{1+15H^{2}\beta+9H^{2}\beta\tau}{4H^{2}\beta\tau}\right)
\tilde{\phi}_{\tilde{k}}=0
\end{equation}
The creation ``time" (when a mode outgrows the minimum length) now
corresponds for all \(\tilde{k}\) to \(\tau_{c}=-1\). (This is particular
to the de Sitter case, where \(H\) is a constant.) The time \(\tau_{c}\) is still a
singular point of (\ref{eofmexplicit}). However, it is now a regular singular
point since the Taylor expansions of the term
\begin{equation*}
\left(\tau-\tau_{c}(\tilde{k})\right)^{2}\cdot\left(\frac{1+15H^{2}\beta+9H^{2}\beta\tau}{4H^{2}\beta\tau}\right)
\end{equation*}
and the term
\begin{equation*}
\left(\tau-\tau_{c}(\tilde{k})\right)\cdot\frac{5+\tau}{2\tau(1+\tau)}
\end{equation*}
indeed exist. For later convenience, we shift the \(\tau\)-variable once more so that
\(x:=\tau+1\); creation time for all modes is now \(x_{c}=0\):
\begin{equation}\label{eofminx}
\tilde{\phi}_{\tilde{k}}''+\frac{4+x}{2x(x-1)} \tilde{\phi}_{\tilde{k}}'
-\left(\frac{1+6H^{2}\beta+9H^{2}\beta x}{4H^{2}\beta(x-1)}\right)
\tilde{\phi}_{\tilde{k}}=0
\end{equation}
Having transformed the creation time into a regular singular point, we are now ready to solve the mode equation through the Frobenius method.

%We will now give a brief overview of the Frobenius method as needed for our present analysis and then use it to solve the equation of motion (\ref{eofminx}).
%Note that this equation was obtained making use of the de Sitter case \(a(\eta)=-1/H\eta\), and the solution presented below will be valid for the de Sitter case only. We comment on the generalization of our approach in Section \ref{generalizedapproach}.

\subsection{The Frobenius solutions}\label{solutiondesitter}
Full details on the Frobenius method are given in Appendix \ref{frobenius}. %In our case, the Frobenius solutions to (\ref{eofminx}) will converge for \(x<1\) (i.e. for all times with the exception of the infinite future). Moreover, the Frobenius indicial equation (\ref{characteristic}) in the Appendix here has the two solutions  \(r_{1}=3\) and \(r_{2}=0\).\\
% Before we derive the explicit solutions to (\ref{eofminx}), let us state that from the Taylor expansions of \((\tau-\tau_{c})\cdot \left(\frac{\tilde{\nu}'}{\tilde{\nu}}+\frac{\left(\frac{\partial}{\partial\tau}\frac{\partial\tau}{\partial\eta}\right)}{\frac{\partial\tau}{\partial\eta}}\right)\) and \((\tau-\tau_{c})^{2}\cdot\left[\frac{\tilde{\mu}}
%{\left(\frac{\partial\tau}{\partial\eta}\right)^{2}}-
%3\left(\frac{\tilde{a}''}{\tilde{a}}\right)-6\left(\frac{\tilde{a}'}{\tilde{a}}\right)^{2}-
%3\frac{\tilde{a}'\tilde{\nu}'}{\tilde{a}\tilde{\nu}}-3\frac{\tilde{a}'}{\tilde{a}}\frac{\left(\frac{\partial}{\partial\tau}\frac{\partial\tau}{\partial\eta}\right)}{\frac{\partial\tau}{\partial\eta}}\right]\) we can already determine the radius of convergence \(\rho\) of the solutions. It turns out that they converge converge for \(x<1\), and consequently so do the Frobenius solutions we are about to determine. Since our variable \(x\) only runs from 0 to 1, they therefore converge for all times with the exception of \(x=1\) which corresponds to the infinite future.
%In our case, the Taylor expansions of the coefficients give the indicial equation
%\begin{equation}\label{firstline}
%r\cdot (r-1)-2r=r\cdot (r-3)=0
%\end{equation}
%with the two solutions \(r_{1}=3\) and \(r_{2}=0\).
Using the Frobenius ansatz (see  %\footnote{Since we will need a general (\(r\)-dependent) expression of \(a_{n}\) for later purposes, we will at first not make use of the already determined values of \(r_{1}\) and \(r_{2}\). Note also that our variable \(x\) runs from 0 to 1, therefore we do not need to take the absolute value.} \(\tilde{\phi}_{\tilde{k}}(x)=x^{r}\sum_{n=0}^{\infty}a_{n}(r)\cdot x^{n}\)
(\ref{sol1}) in the Appendix) we obtain from (\ref{eofminx}):
\begin{eqnarray}
&&\sum_{n=0}^{\infty}a_{n}(n+r)(3-n-r)\cdot x^{n-1}\nonumber\\
&+&\sum_{n=0}^{\infty}a_{n}(n+r)(n+r-1/2)\cdot x^{n}\nonumber\\
&-&\sum_{n=0}^{\infty}a_{n}\left(\frac{1}{4H^{2}\beta}+3/2\right)\cdot x^{n+1}\nonumber\\
&-&9/4\cdot\sum_{n=0}^{\infty}a_{n}\cdot x^{n+2}=0\label{inserted}
\end{eqnarray}
%Since the coefficient in front of every order of \(x\) has to vanish separately,
The first line of (\ref{inserted}) yields the indicial equation
(\ref{characteristic}), which here has the solutions \(r_{1}=3\) and \(r_{2}=0\). The first coefficient \(a_{0}\) is an
arbitrary normalization constant and we set \(a_{0}=1\). Comparing the terms of equal powers in  \(x\), we then find
\begin{eqnarray*}
a_{1}&=&\frac{r(2r-1)}{2(r+1)(r-2)},\\
a_{2}&=&\frac{1}{4(2+r)(1-r)}\\
&&\times\left(\frac{1}{H^{2}\beta}+6-\frac{r(2r-1)(2r+1)}{r-2}\right).
\end{eqnarray*}
From \(x^{2}\) on, all terms in (\ref{inserted}) contribute to the
coefficient, thus we can find a recursion formula for \(a_{n}\) for
\(n\geq 3\) as a function of \(a_{n-1}, a_{n-2}\) and \(a_{n-3}\) as well
as \(r\):
\begin{eqnarray}
a_{n}(r)&=&\frac{1}{(n+r)(n+r-3)}\label{recursion}\\
&&\times\bigg(-\frac{9}{4}a_{n-3}-\frac{1}{4}a_{n-2}\left[\frac{1}{H^{2}\beta}+6\right]\nonumber\\
&&+\frac{1}{2}a_{n-1}(n+r-1)(2n+2r-3)\bigg)\nonumber
\end{eqnarray}
%For the first solution we have \(r=r_{1}=3\) and therefore
%\begin{eqnarray}
%a_{1}&=&\frac{15}{8}\cdot a_{0}\\
%a_{2}&=&-\frac{1}{40}\left(\frac{1}{H^{2}\beta}-99\right)\cdot a_{0}.
%\end{eqnarray}
%We have found the first solution (\(r_{1}=3\)) to the equation (\ref{eofminx}) to be given by
%\begin{equation}\label{first}
%\tilde{\phi}_{\tilde{k},1}(x)=x^{3}\sum_{n=0}^{\infty}a_{n}(r=3)\cdot x^{n},
%\end{equation}
%where the coefficient \(a_{0}\) can be chosen freely, e.g. \(a_{0}=1\). The \(a_{n}\) for \(n\geq 3\) follow from (\ref{recursion}).
The first solution to (\ref{eofminx}) is simply given by inserting \(r=r_{1}=3\) in the above expressions.\\
Since \(r_{1}-r_{2}=N=3\) is a positive integer, the solution for
\(r_{2}=0\) is of the form given by (\ref{sol2}) in the Appendix. %We can make use of our result for the first solution (\ref{first}) to determine the missing constant \(A\) and the coefficients \(c_{n}\).
The constant \(A\) can be calculated from \(a_{3}\) as
\begin{equation}
A%&=&\lim_{r\rightarrow r_{2}}(r-r_{2})\cdot a_{N}(r)\nonumber\\
=\lim_{r\rightarrow 0}r\cdot a_{3}(r)=\frac{1}{8H^{2}\beta}\cdot
a_{0}\label{A}.
\end{equation}
The coefficients \(c_{n}(r=0)\) for \(n=0,1,2,\dots\)
are related to the \(a_{n}\) in (\ref{recursion})  by%\footnote{Here it becomes clear why we calculated the general expression for \(a_{n}(r)\) - we need to take the \(r\)-derivative to find the formula for the \(c_{n}\).}
\begin{equation}
c_{n}(r_{2}=0)=\left[\frac{d}{dr}\left((r-r_{2})a_{n}(r)\right)\right]_{r=r_{2}=0}.
%=\left[\frac{d}{dr}\left(ra_{n}(r)\right)\right]_{r=0}\label{cn}.
\end{equation}
%This gives us
%\begin{eqnarray}
%c_{1}&=&0,\\
%c_{2}&=&\frac{1}{8}\left(\frac{1}{H^2\beta}+6\right)\cdot a_{0}.
%\end{eqnarray}
%(To check whether the values for \(A\) and \(c_{n}\) found above are correct, we can also calculate the second solution directly by inserting the ansatz (\ref{sol2}) into the equation (\ref{eofminx}).) The second solution to (\ref{eofminx}) therefore reads
%\begin{equation}\label{second}
%\tilde{\phi}_{\tilde{k},2}(x)=\frac{1}{8H^{2}\beta}\cdot x^{3}\sum_{n=0}^{\infty}a_{n}(r=3)x^{n} \cdot\ln x+ \sum_{n=0}^{\infty}c_{n}(r=0)x^{n}.
%\end{equation}
%Here, the \(c_{n}\) are given by (\ref{cn}).
The second solution to (\ref{eofminx}) is then obtained by setting
\(r=r_{2}=0\). (Equivalently, \(A\) and the coefficients \(c_{0}\) can be
calculated directly by inserting (\ref{sol2}) into (\ref{inserted}), see
Appendix \ref{frobenius}.) The two independent solutions can be combined
to form the exact general solution of (\ref{eofminx}),
%(\ref{first}) and (\ref{second}) are two independent solutions for the
%equation of motion (\ref{eofminx}). The general solution is therefore given
%by their linear combination:
\begin{equation}\label{general}
\tilde{\phi}_{\tilde{k}}(x)=C_{1}\tilde{\phi}_{\tilde{k},1}(x)+C_{2}\tilde{\phi}_{\tilde{k},2}(x).
\end{equation}
It is clear from Frobenius theory that the power series solutions
converge for all $x<1$,, i.e. all \(\tau<0\). Here, \(C_{1}\) and \(C_{2}\) are complex constants whose
choice corresponds to the choice of vacuum.
Note that our solutions (\ref{general}) could be rewritten
in terms of the conformal time \(\eta\) and would then of course solve equation
(\ref{equationofmotion}).

\subsection{Wronskian condition}\label{wronskiancondition}
%\subsubsection*{Transformed equation of motion}
After the transformation
\(\eta\rightarrow\tau\) the Wronskian condition
(\ref{wronskian}) takes the form:
\begin{equation}\label{newwronskian}
\tilde{\nu}(\tau,\tilde{k})\frac{1}{\eta'(\tau,\tilde{k})}\left[\tilde{\phi}_{\tilde{k}}(\tau)\tilde{\phi}'^{*}_{\tilde{k}}(\tau)-
\tilde{\phi}^{*}_{\tilde{k}}(\tau)
\tilde{\phi}_{\tilde{k}}'(\tau)\right]=i
\end{equation}
Note that while the solution space to the mode equation has two complex and therefore four real dimensions, the Wronskian condition reduces the
dimensionality to three real dimensions (since only the imaginary part of (\ref{newwronskian}) is nontrivial).

For example, in the case of de Sitter space and when splitting the constants \(C_{1}\) and \(C_{2}\) in (\ref{general}) into
their real and imaginary parts, \(C_{j}=R_{j}+iK_{j},\,j=1,2\), the Wronskian condition at the creation time \(\tau_{c}=-1\) can be expressed as:
\begin{equation}
R_{1}K_{2}-K_{1}R_{2}=\left(\frac{12H}{\beta^{3/2}\tilde{k}^{3}}\right)^{-1}.
\end{equation}
%We have presented explicit Frobenius series solutions to the equation of motion (\ref{eofmexplicit}) and its frictionless analogue.

\subsection{The generic case}\label{generalizedapproach}
We have here used a suitable transformation of the time variable in order to turn the creation time into a regular singular point so that then the Frobenius method yielded the exact solutions to the mode equation in de Sitter space. In fact, this method, using in all cases the same transformation from conformal time to the $\tau$ variable, can also be applied to more general FRW spacetimes. In particular, we have applied the method to the case of general power-law backgrounds. The exact solutions are then only slightly more difficult to calculate, see Appendix \ref{powerlaw}. Namely, while in the de Sitter case we obtained a recursion formula for the
coefficients \(a_{n},c_{n}\) as a function of a limited number of
predecessors (namely three), in a general power-law background each
coefficient generally depends on all others that precede it.
Interestingly, the roots of the indicial equation in the general power-law case turn out to be the same
as for the de Sitter case, namely \(r_{1}=3, r_{2}=0\). Therefore the behaviour of a mode close to its creation time is qualitatively the same in both the de Sitter and the general power-law case.

Further, it is often useful to employ field redefinitions, for example in order to eliminate friction terms in the mode equation. Our method  remains applicable under such field redefinitions. In particular, as shown in \cite{kempfniemeyer01}, the field redefinition \(\psi_{\tilde{k}}(\eta)=\sqrt{\nu}\phi_{\tilde{k}}(\eta)\) eliminates the friction term in (\ref{equationofmotion}), so that we arrive at an equation of the form \(\psi_{\tilde{k}}''+\omega_{\tilde{k}}^{2}(\eta)\psi_{\tilde{k}}=0\).
Similarly, it is possible to eliminate the friction term in our mode equation (\ref{eofminx}) in which the creation time is a regular singular point. Thereby, the creation time remains a regular singular point.
The corresponding field redefinition and exact solutions are presented in Appendix \ref{frictionless}.

Note that formulating the mode equation in frictionless form allows us not
only to obtain the exact solutions through the Frobenius method but also
to obtain a second set of solutions using the WKB method in the adiabatic
regime. This will provide us with two bases of the solution space. These
are of course related
by a Bogolyubov transformations and we will calculate it
 Sec.\ref{criteria}.\\%A similar transformation is also possible after the transformation to the \(\tau\) variable, and the resulting equation can then also be solved by the Frobenius method (see Appendix \ref{frictionless}).\\

%We will not give the Frobenius solutions to other formulations of the equation of motion here, but comment on other cases to which we have %applied this formalism. It has already been mentioned that our field \(\phi_{\tilde{k}}(\eta)\) differs from the ``physical'' variable, the %intrinsic curvature \(\Phi_{\tilde{k}}(\eta)\), by a factor of \(a^{-3}(\eta)\): \(\Phi_{\tilde{k}}(\eta)= %a^{-3}(\eta)\phi_{\tilde{k}}(\eta)\). %(This was, e.g., also the field considered in \cite{amjad}.)
%If we rewrite (\ref{equationofmotion}) as an equation for
%\(\Phi_{\tilde{k}}(\eta)\), it can equally be transformed to the
%\(\tau\)-variable and then solved with the Frobenius method.

\section{Initial conditions}\label{poseinitial}
%We have shown that for late times \(\eta\rightarrow0\), the modified mode equation with an UV cutoff leads back to the usual mode equations and the exact solutions must therefore show the same behaviour as the well-known solutions for the usual mode equations.
%We will now turn our attention to early times close to the creation time \(\eta_{c}\), when the difference between the usual and the modified mode equation is significant.
Having derived the space of exact solutions to the exact mode equation, it becomes crucial
to identify the physical solution in that space, since this then implicitly identifies the vacuum state.

For this purpose, given that the mode equation is of second order, it would appear to suffice to specify suitable initial conditions at the mode's creation time, such as the physical solution's value and derivative. Interestingly, the behaviour of the exact solutions towards the creation time reveals that the situation is more complex.

Indeed, the values of \(\phi_{\tilde{k}}(\eta)\) and
its derivative with respect to conformal time, at creation time are given by:
\begin{eqnarray}
\phi_{\tilde{k}}(\eta_{c})&=&C_{2}\label{functioninetaatcreation}\\
\phi'_{\tilde{k}}(\eta_{c})&=&-\frac{1}{2\eta_{c}}~C_{2}\left(6+\frac{1}{H^{2}\beta}\right)\label{derivinetaatcreation}
\end{eqnarray}
This shows that prescribing $\phi_{\tilde{k}}(\eta_{c})$ and $\phi'_{\tilde{k}}(\eta_{c})$ will fix only $C_2$ but not $C_1$.
In fact, at the creation time, it is also not possible to specify a solution by specifying higher order derivatives of $\phi_{\tilde{k}}$ since, as one readily verifies, the behaviour of the exact solutions shows that all these higher derivatives are necessarily divergent. For example, \(\phi''_{\tilde{k}}(\eta=\eta_{c})\) diverges as \(\propto\frac{1}{\left(\textnormal{plog}(-\beta\tilde{k}^{2}/a^{2})+1\right)}\).
Roughly speaking, as we approach creation time, the general solution (\ref{general}) loses its dependence on \(C_{1}\).

%Since we cannot specify initial conditions in the usual way by giving the function's and its derivative's value at creation time (the earliest time for which our mode equation (\ref{equationofmotion}) is valid),
While the missing $C_1$ has two real dimensions, one of these is fixed by
the Wronskian condition. Within the remaining one-parameter family of
solutions, a solution can be picked by choosing a phase condition, for
example, by imposing that the solution be real at a given time
\(\eta_{aux}\neq\eta_{c}\), i.e.,
\(Im\left(\phi_{\tilde{k}}(\eta_{aux})\right)=0\).

%Since both of the solutions \(\phi_{\tilde{k},1}(\eta)\) and \(\phi_{\tilde{k},2}(\eta)\) are real, the imaginary part of (\ref{general}) is simply given by \(\Im\left(\phi_{\tilde{k}}(\eta)\right)=K_{1}\phi_{\tilde{k},1}(\eta)+K_{2}\phi_{\tilde{k},2}(\eta)\). (\ref{imphi}) then leads to
%\begin{equation}\label{K1fromphase}
%K_{1}=-K_{2}\left(A\ln\left(\textnormal{plog}(-\beta\tilde{k}^{2}/a^{2})+1\right)+\frac{\sum_{n=0}^{\infty}c_{n}\left(\textnormal{plog}(-\beta\tilde{k}^{2}/a^{2})+1\right)^{n}}{\sum_{n=0}^{\infty}a_{n}\left(\textnormal{plog}(-\beta\tilde{k}^{2}/a^{2})+1\right)^{n+3}}\right).
%\end{equation}
%\(R_{1}\) follows from the Wronskian condition (\ref{wronskian}).

Thus, any mode function can be specified by  giving the mode function's value at creation
time \(\eta_{c}\) (two real parameters) and demanding that the solution be
real at an auxiliary time \(\eta_{aux}\) (one parameter). The Wronskian
condition constrains the fourth parameter.

\section{Comparison with the results of Easther \emph{et al.}}\label{criteria}
%\subsection{Approximate solution}
In earlier work, see \cite{greeneetal0104}, the irregular singular point of the mode equation
(\ref{equationofmotion}) was dealt with by truncating a series expansion of the coefficients of the mode equation. For the resulting approximate mode equation\footnote{In \cite{greeneetal0104} a different notation for this
equation was chosen; apart from replacing the field \(\phi_{k}\) by
\(u_{k}=\phi_{k}/a^{2}\), also the variable \(y\) was introduced by setting \(\eta=\eta_{c}(1-y)\), and all expressions were re-written in terms of \(y\).}
\begin{equation}\label{approximateeofm}
\phi_{\tilde{k}, approx}''-\frac{1}{2(\eta-\eta_{c})}\phi_{\tilde{k},
approx}'+\frac{\mathscr{A}}{(\eta-\eta_{c})}\phi_{\tilde{k}, approx}=0,
\end{equation}
\(\eta=\eta_{c}\) is a regular singular point. Here, \(\mathscr{A}=\frac{\tilde{k}\sqrt{e}}{4\sqrt{\beta} H}(1+6\beta
H^{2})\). The solution space of the approximate mode equation (\ref{approximateeofm}) was found to be spanned by
\begin{equation}\label{F}
F(\eta)=\left(\frac{\sqrt{\mathscr{A}}}{2}+i\mathscr{A}\sqrt{\eta-\eta_{c}}\right)\exp(-2i\sqrt{\mathscr{A}(\eta-\eta_{c})})
\end{equation}
and its complex conjugate, so that the general solution is given by the
linear combination:
\begin{equation}
\phi_{\tilde{k},approx}=\mathscr{C}_{1}F(\eta)+\mathscr{C}_{2}F^{*}(\eta)
\end{equation}
Since we have found in Sec.\ref{solutiondesitter} the exact solutions (\ref{general}) to the exact mode
equation (\ref{equationofmotion}), we can now compare them with the
solutions to the approximate mode equation that were obtained in \cite{greeneetal0104}.

To compare the
behaviour of the exact and the approximate solution close to the creation
time, we take the first and second \(\eta\) derivatives of \(F(\eta)\)
%\begin{eqnarray}
%F'(\eta)&=&A^{3/2}\exp\left(-2i\sqrt{A(\eta-\eta_{c})}\right)\label{Fprime}\\
%F''(\eta)&=&-i\frac{A^{2}}{\sqrt{\eta-\eta_{c}}}\exp\left(-2i\sqrt{A(\eta-\eta_{c})}\right)\label{F2prime}
%\end{eqnarray}
%Evaluating (\ref{F}) and (\ref{Fprime}) at creation time, we find for the approximate solution \(\phi_{\tilde{k}, approx}(\eta)\) and its first derivative
and evaluate at creation time, which gives
\begin{eqnarray}
\phi_{\tilde{k}, approx}(\eta=\eta_{c})&=&\frac{\sqrt{\mathscr{A}}}{2}(\mathscr{C}_{1}+\mathscr{C}_{2}),\\
\phi'_{\tilde{k},
approx}(\eta=\eta_{c})&=&\mathscr{A}^{3/2}(\mathscr{C}_{1}+\mathscr{C}_{2}).
\end{eqnarray}
We encounter the same behaviour we observed in the case of the exact
solution \(\phi_{\tilde{k}}(\eta)\): the function and its first derivative
become proportional to another for $\eta\rightarrow \eta_c$ and the proportionality factor is the same as in (\ref{functioninetaatcreation})
and (\ref{derivinetaatcreation}), namely
\(2\mathscr{A}=\frac{\tilde{k}\sqrt{e}}{2\sqrt{\beta} H}(1+6\beta
H^{2})\). Thus, also the solutions to the approximate mode equation show that posing initial conditions on $\phi$ and its derivative at $\eta_c$ does not suffice to specify a solution.

Further, the second derivative of the solutions to the approximate mode equation
diverges for \(\eta\rightarrow\eta_{c}\), namely as
\(F''(\eta)\propto\frac{1}{\sqrt{\eta-\eta_{c}}}\). In comparison, the second
\(\eta\)-derivative of the exact solution \(\phi_{\tilde{k}}(\eta)\) at
creation time diverges as
\(\propto\left(\textnormal{plog}(-\beta\tilde{k}^{2}H^{2}\eta^{2})+1\right)^{-1}=\frac{\sqrt{-\eta_{c}}}{2}\frac{1}{\sqrt{\eta-\eta_{c}}}+\mathscr{O}(1)\)
(using the series expansion of the plog function). Thus, the
solutions to the approximate mode equation also properly reproduce the leading divergence for $\eta\rightarrow\eta_c$.

%This analysis makes it clear that we have to be aware of the limited range of \(\eta\) over which the approximate solution should be used: Only very close to creation time \(\eta_{c}\) will it correctly reflect the behaviour of the exact solution. Already in their second derivatives, the exact and the approximate solution differ.
We note that the next to leading orders are different: the approximate
equation has a regular singular point at creation time which meant that
the Frobenius method could be used to derive two power series solutions.
The roots of the indicial equation are $(0, 3/2)$, unlike the roots
($0,3$) of the exact mode equation. This means that the solutions to the
approximate mode equation are of the form:
 \(\phi_{\tilde{k},approx,1}=\sum
a_{n}(\eta-\eta_{c})^{n}\) and \(\phi_{\tilde{k},approx,2}=\sum
c_{n}(\eta-\eta_{c})^{n+3/2}\) The absence of the logarithm in these
functions implies that their higher derivatives exhibit a different
divergent behaviour than that of the solutions to the exact mode equation.

Related to this, it is clear that the approximate mode equation does not
conserve the Wronskian of the exact mode equation. Instead, it conserves
the expression \(|\mathscr{C}_{1}|^{2}-|\mathscr{C}_{2}|^{2}\). Indeed,
the Wronskian of the exact mode equation, (\ref{wronskian}), for the
solutions to the approximate mode equation is given by
\begin{equation}\label{w-approx}\mathscr{W}(\eta) =
2\,\nu(\eta,\tilde{k})\,\mathscr{A}^{5/2}\sqrt{\eta-\eta_{c}}
\left(|\mathscr{C}_{1}|^{2}-|\mathscr{C}_{2}|^{2}\right),
\end{equation}
where \(\nu(\eta,\tilde{k})\), given by (\ref{nu}), has a nontrivial time
dependence. Using again the series expansion around \(\eta_{c}\), we find
that: $$
\frac{1}{\textnormal{plog}(-\beta\tilde{k}^{2}H^{2}\eta^{2})+1}=\frac{\sqrt{-\eta_{c}}}{2}
\frac{1}{\sqrt{\eta-\eta_{c}}}+\mathscr{O}(1)
$$
Thus, at \(\eta_{c}\), it is possible to enforce the exact Wronskian
condition on the solutions to the approximate mode equation, though the
Wronskian will be conserved only to first order in $\sqrt{\eta-\eta_{c}}$.
This shows that if a solution to the approximate mode equation is to be
matched to the numerical evolution of the exact mode equation, as was done
in \cite{greeneetal0104}, then this match needs to be performed at a time
$\eta_{aux}$ which is as close as possible to the creation time. We note,
however, that as our exact solutions will show below, there is limit to
how close $\eta_{aux}$ can be chosen to the creation time, because of
unavoidable numerical instabilities.

%\subsection{Match between exact and approximate solution}
In \cite{greeneetal0104}, a particular choice of initial conditions was
suggested, based on a formal similarity between certain solutions to the
approximate equation and the solutions that correspond to the Bunch-Davies
vacuum in inflation without a cutoff. The argument is that, if the
``friction'' term in the approximate mode equation (\ref{approximateeofm})
could be ignored,
%\begin{equation}\label{frictionignored}
%\phi_{\tilde{k}, approx}''+\frac{A}{(\eta-\eta_{c})}\phi_{\tilde{k}, approx}=0,
%\end{equation}
then it would be of the form \(\phi_{k}''+\omega_{k}^{2}(\eta)\phi_{k}=0\) with \(\omega_{k}=\sqrt{\frac{\mathscr{A}}{(\eta-\eta_{c})}}\). In the region where the adiabaticity condition \(\left|\frac{\omega'_{\tilde{k}}(\eta)}{\omega_{\tilde{k}}^{2}(\eta)}\right|\ll 1\) is satisfied, this equation has two approximate solutions of the WKB form:%, \(\phi_{k}^{\mp}(\eta)=\frac{1}{\sqrt{2\omega_{k}}}\exp\left(\pm i\int_{\eta_{i}}^{\eta}\omega_{k}(\eta')d\eta'\right)\); analogously, Easther \emph{et al.} state that this equation then would have a set of solutions
\begin{eqnarray}
\phi_{\tilde{k},approx}^{\mp}(\eta)%&=&\frac{1}{\sqrt{2\omega_{\tilde{k}}}}\exp\left(\pm i\int_{\eta_{i}}^{\eta}\omega_{\tilde{k}}(\eta')d\eta'\right)\nonumber\\
&=&\left(\frac{\eta-\eta_{c}}{4\mathscr{A}}\right)^{1/4}\nonumber\\
&&\times\exp\left(\pm2i\sqrt{\mathscr{A}(\eta-\eta_{c})}\right).\label{phifrictionignored}
\end{eqnarray}
%In the standard inflationary scenario without a cutoff, the vacuum is usually chosen to be the so-called Bunch-Davies vacuum which reduces to the Minkowski vacuum for wavelength much shorter than the Hubble scale. The Bunch-Davies vacuum amounts to choosing only the solution that stands in front of the positive frequency modes in the field operator \(\hat{\phi}(\eta,\vec{x})\), which would be \(\phi_{\tilde{k},approx}^{-}\) in (\ref{phifrictionignored}).\\
Even though one cannot actually ignore the ``friction'' term in the approximate mode
equation because the period close to $\eta_c$ is not adiabatic, it is suggestive that these would-be WKB type functions
possess the same oscillatory behaviour as Easther \emph{et al.}'s solution to their approximate mode equation.

In formal analogy to choosing the Bunch-Davies vacuum solution, Easther \emph{et al.} therefore choose \(\mathscr{C}_{2}=0\) so that their preferred physical solution to their approximate mode equation
reads\footnote{Note that even though this solution resembles that of an
adiabatic vacuum, it is not because the adiabaticity condition with
\(\omega_{k}=\sqrt{\frac{\mathscr{A}}{(\eta-\eta_{c})}}\) is not
satisfied.}
\begin{eqnarray*}
\phi_{\tilde{k},approx}&=&\mathscr{C}_{1}\left(\frac{\sqrt{\mathscr{A}}}{2}+i\mathscr{A}\sqrt{\eta-\eta_{c}}\right)\\
&&\times\exp(-2i\sqrt{\mathscr{A}(\eta-\eta_{c})}).
\end{eqnarray*}
The real and imaginary part of \(\mathscr{C}_{1}\) are related via the Wronskian condition. The remaining arbitrariness is merely the freedom to choose an overall phase, and can be fixed by demanding, for example, that $\phi_{\tilde{k},approx}$ is real at some auxiliary time \(\eta_{aux}\).

Let us now investigate which exact solution to the exact mode equation
Easther \emph{et al.}'s choice corresponds to. Normally, in the case of
second order wave equations, in order to match two solutions at any
arbitrary point it is sufficient to set these solutions and their first
derivatives equal at that point. We would obviously like to match Easther
\emph{et al.}'s function to one of our exact solutions to the exact
equation at the creation time. However, such a match-up at creation time
itself does here not suffice to use Easther \it et al.\rm 's choice of
mode function to uniquely pick out an exact solution to the exact mode
equation. This is because, as we saw earlier, at the creation time
\(\eta=\eta_{c}\), the value and the first derivative of the solutions are
not independent (but are instead a fixed multiple of another).

We therefore studied matching-up the mode function chosen by Easther
\emph{et al.} with an exact solution by setting the functions and their
derivatives equal at times \(\eta_{m}>\eta_{c}\). We were indeed able to
reproduce the observation made in \cite{greeneetal0104} that at times
considerably later than $\eta_c$ the mode with \(\mathscr{C}_{2}=0\) shows
close to adiabatic evolution with small characteristic oscillations. This
match-up is very delicate, however, and we can now see the underlying
reasons.

First, the match cannot be improved by choosing earlier and earlier match-up times  $\eta_m$ because the mode function's amplitude and  derivative loose their independence for $\eta_m\rightarrow \eta_c$, leading to numerical instability of the match-up.

Second, the match also cannot be improved by choosing late matching times $\eta_m$   because the simplified mode equation obeyed by
\(\phi_{\tilde{k},approx}(\eta)\) is close to the precise mode equation only for early $\eta$. Indeed, recall that the function chosen by Easther \emph{et al.} conserves the Wronskian of the exact mode equation only very close to $\eta_c$.

Given that there is no ideal match-up procedure, one may alternatively follow a strategy already outlined in Sec.\ref{poseinitial}: the amplitudes of the two solutions are matched at creation time, which fixes $C_2$. The information about $C_1$ is then obtained by enforcing the Wronskian condition and by choosing the overall phase such that the mode function is, for example, real at some auxiliary time. %To fix the phase, we can demand that the exact and the approximate solution should be in phase at some auxiliary time \(\eta_{aux}\).
We observed again a notable dependence on the arbitrary auxiliary time
\(\eta_{aux}\): it must be chosen not too late after creation time because
the approximate mode equation that the function of Easther \emph{et al.}
obeys is close to the exact mode equation only in the vicinity of
\(\eta_{c}\), but neither must $\eta_{aux}$ be chosen too early because
close to \(\eta_{c}\) the exact solution is entirely dominated by one of
its dimensions in the solution space and information about the
contribution of the other dimension is numerically increasingly difficult
to extract.

\section{Comparison with the WKB solution}\label{WKB}
The starting point of our present analysis was the mode equation (\ref{equationofmotion}), which is analogous to that of a damped harmonic oscillator.
%As was shown in \cite{kempfniemeyer01}, a suitable field redefinition renders (\ref{equationofmotion}) frictionless.
We expressed (\ref{equationofmotion}) in the \(\tau\)-variable to obtain (\ref{eofmexplicit}), which we were then able to solve by means of the Frobenius method. As we show in Appendix \ref{frictionless}, a suitable field redefinition brings (\ref{eofmexplicit}) into a frictionless form,
\begin{equation}
\tilde{\psi}_{\tilde{k}}''+\tilde{\omega}^{2}(\tau)\tilde{\psi}_{\tilde{k}}=0,
\end{equation}
see (\ref{trafowithouteofm}), where
\begin{eqnarray}
\tilde{\omega}^{2}(\tau)&=&-\frac{1}{16{H}^{2}\beta\,{\tau}^{2}
\left (\tau+1\right )^{2}}\times\bigg(36\,\beta\,{\tau}^{4}{H}^{2}\nonumber\\
&&+132\,\beta\,{\tau}^{3}{H}^
{2}+153\,\beta\,{\tau}^{2}{H}^{2}+30\,\beta\,\tau\,{H}^{2}\nonumber\\
&&+5\,\beta\,{H}^{2}+4\,{\tau}^{3}+8\,{\tau}^{2}+4\,\tau\bigg).\label{omegaadiab}
\end{eqnarray}
Also this mode equation can be solved with the Frobenius method. We write the general solution in the form
\begin{equation}
\label{ab}
\tilde{\psi}_{\tilde{k}}(\tau)=D_{1}\tilde{\psi}_{\tilde{k},1}(\tau)+D_{2}\tilde{\psi}_{\tilde{k},2}(\tau),
\end{equation}
see (\ref{generalwithout}) in Appendix \ref{frictionless}.

The advantage of the frictionless formulation is that
in the adiabatic range of \(\tau\), i.e,  where
\(\left|\frac{\tilde{\omega}'(\tau)}{\tilde{\omega}^{2}(\tau)}\right|\ll
1\) is fulfilled, (\ref{trafowithouteofm}) will have a set of solutions of
the WKB form, namely
\begin{equation}
\tilde{\Psi}_{\tilde{k}}^{\pm}(\tau)=\frac{1}{\sqrt{2\tilde{\omega}(
\tau)}}\exp\left(\mp{}i\int_{\tau_{i}}^{\tau}\tilde{\omega}(\tau')d\tau'\right).\label{plus}
\end{equation}
A general solution to (\ref{trafowithouteofm}) in the adiabatic regime is
then given by
\(\tilde{\Psi}_{\tilde{k}}(\tau)=\lambda_{-}\tilde{\Psi}_{\tilde{k}}^{-}(\tau)+\lambda_{+}\tilde{\Psi}_{\tilde{k}}^{+}(\tau)\).
Note that the Wronskian condition, when evaluated at \(\tau_{c}=-1\), imposes:
\begin{equation}
|\lambda_{+}|^{2}-|\lambda_{-}|^{2}=\frac{\tilde{k}^{3}\beta^{3/2}}{2H}
\end{equation}
In the adiabatic regime, the exact solutions and the WKB solutions must be related through
a Bogolyubov transformation.
The WKB solutions, (\ref{plus}), suggest a special choice of the vacuum state, namely
the adiabatic vacuum characterized by \(\lambda_{-}=0\). The Bogolyubov
transformation will then tell us which combination of the exact solutions
\(\tilde{\psi}_{\tilde{k},1}(\tau)\) and
\(\tilde{\psi}_{\tilde{k},2}(\tau)\) this choice corresponds to, i.e., it will determine the coefficients
\(D_{1},D_{2}\) in (\ref{ab}).

For the purpose of plotting the properties of the exact solutions, their Frobenius power series expansion,
 (\ref{generalwithout}), needs to be truncated at some finite order \(N\). We established that
choosing any  \(N>30\) suffices to ensure that the plot of the truncated function is valid well into the adiabatic regime.

In order to find the Bogolyubov transformation between the two
sets of solutions, it is sufficient to match the value and derivative of the adiabatic solution to an exact solution at some arbitrary time \(\tau_{ad}\) in the adiabatic regime:
\begin{eqnarray*}
&&D_{1}\tilde{\psi}_{\tilde{k},1}(\tau_{ad})+D_{2}\tilde{\psi}_{\tilde{k},2}(\tau_{ad})\\
&&\qquad=\lambda_{-}\tilde{\Psi}_{\tilde{k}}^{-}(\tau_{ad})+\lambda_{+}\tilde{\Psi}_{\tilde{k}}^{+}(\tau_{ad})\label{match1}\\
&&D_{1}\tilde{\psi}_{\tilde{k},1}'(\tau_{ad})+D_{2}\tilde{\psi}_{\tilde{k},2}'(\tau_{ad})\\
&&\qquad=\lambda_{-}\tilde{\Psi}_{\tilde{k}}'{}^{-}(\tau_{ad})+\lambda_{+}\tilde{\Psi}_{\tilde{k}}'{}^{+}(\tau_{ad})\label{match2}
\end{eqnarray*}
We verified that the result of the match-up is indeed independent of the match-up time within the adiabatic regime.

\subsubsection*{Adiabatic vacuum}
We are particularly interested in that mode function
\(\tilde{\psi}_{\tilde{k}}(\tau)\) which corresponds to the adiabatic
vacuum (to zeroth order) in the adiabatic regime. This choice of vacuum is defined by \(\lambda_{-}=0\) since in the
adiabatic limit only the modes of positive frequency are present. From the
match-up of our two sets of solutions we then have
\begin{eqnarray*}
D_{1}&=&-\frac{1}{3}\left(\tilde{\Psi}_{\tilde{k}}^{+}\tilde{\psi}'_{\tilde{k},2}-\tilde{\psi}_{\tilde{k},2}\tilde{\Psi}'{}_{\tilde{k}}^{+}\right)\cdot\lambda_{+},\\
D_{2}&=&\tilde{\psi}'{}_{\tilde{k},2}^{-1}\cdot\left(\lambda_{+}\tilde{\Psi}'{}_{\tilde{k}}^{+}-D_{1}\tilde{\psi}'_{\tilde{k},1}\right).
\end{eqnarray*}
(Here, the fields are understood to be evaluated at the arbitrary time
\(\tau_{ad}\).) The resulting values of $D_1$ and $D_2$ are several orders
of magnitude apart and not obviously special, in the sense that they do
not suggest a particular mathematical criterion that would single out this
initial condition and therefore the vacuum in a way that would not rely on
the adiabatic expansion. We noticed however, that numerically, $D_1$ and
$D_2$ appear to be at close to 90 degrees when viewed as vectors in the
complex plane. In the Outlook we will comment on the potential
significance for the choice of vacuum of the existence of an orthonormal
basis that is canonical in the solution space.

\subsubsection*{Deviations from the adiabatic vacuum}
%In the previous Section, we examined the Bogolyubov transformation that leads from the adiabatic vacuum choice \(\lambda_{-}\equiv0\) to the corresponding combination of the Frobenius solutions \(\tilde{\psi}_{\tilde{k},1}(\tau)\) and  \(\tilde{\psi}_{\tilde{k},2}(\tau)\).
We can now turn this line of argument around and use the knowledge of the exact solutions to bridge the gap between the initial behaviour right at the creation time and the behaviour of the modes in the adiabatic regime. This means that we can precisely link the initial behaviour to the one at horizon crossing which then determines the size and shape of potentially observable effects in the CMB.

Concretely, starting with a certain linear combination of the Frobenius solutions (that may be set by quantum gravity), how close to the adiabatic vacuum will this particular solution be during the adiabatic phase? To this end, we will derive an expression for \(\lambda_{-}\), which measures the deviation from the positive frequency adiabatic solution as a function of \(D_{1}\), \(D_{2}\). %\footnote{\(\lambda_{+}\) can also expressed in this way, but since we are only interested in the deviation from the adiabatic vacuum, we do not need it for now.},
%then derive constraints on \(D_{1},D_{2}\) so that we have only two free
%parameters left.
%\(|\lambda_{-}|^{2}\) as a function of these two
%parameters can then be considered as the ``surface of initial
%conditions'', on which the adiabatic vacuum \(\lambda_{-}\equiv0\)
%corresponds to the miminum.
From the match-up (\ref{match1}) we find
\begin{eqnarray}
\lambda_{-}&=&-i\bigg(D_{1}\left(\tilde{\psi}_{\tilde{k},1}'\tilde{\Psi}_{\tilde{k}}^{+}-\tilde{\Psi}_{\tilde{k}}^{+}{}'\tilde{\psi}_{\tilde{k},1}\right)\nonumber\\
&+&D_{2}\left(\tilde{\psi}_{\tilde{k},2}'\tilde{\Psi}_{\tilde{k}}^{+}-\tilde{\Psi}_{\tilde{k}}^{+}{}'\tilde{\psi}_{\tilde{k},2}\right)\bigg).\label{lambda}
\end{eqnarray}
So far, $\lambda_-$ depends on the real and imaginary parts of  \(D_{1}\) and \(D_{2}\), i.e. on four real parameters:
\begin{eqnarray*}
D_{1}&=&S_{1}+iL_{1}\\
D_{2}&=&S_{2}+iL_{2}
\end{eqnarray*}
Only two of these parameters are independent, however, due to the Wronskian condition
\begin{equation}
S_{1}L_{2}-L_{1}S_{2}=\left(\frac{12H}{\beta^{3/2}\tilde{k}^{3}}\right)^{-1},
\end{equation}
see Appendix \ref{frictionless}, and due to the arbitrariness of the
overall phase, which allows us to choose for example $D_{1}$ real.
$|\lambda_-|^2$ as a function of these remaining two parameters measures
the non-adiabaticity of the vacuum in terms of the initial behaviour of
the physical mode function. Its minimum is zero and denotes the adiabatic
vacuum.

\section{Behaviour of physical quantities near creation time}\label{principles}
So far, we have parameterized the initial behaviour of modes and used our exact solutions to link it to the mode's behaviour at horizon crossing. In order to better understand what might determine the initial behaviour of modes let us now use our exact solutions to investigate the behaviour of physical quantities such as the mode's field uncertainties and its Hamiltonian close to the mode's creation time.

\subsection{Breakdown of the particle picture}
The Hamiltonian expressed through creation and annihilation operators was
given in (\ref{quantizedHamiltonian}).
%Let us quantify how close
%this Hamiltonian is to being diagonal in the Fock basis at some arbitrary time $\eta$.
We see that the nondiagonal terms of the Hamiltonian, i.e., the terms
\(a_{\tilde{k}}a_{-\tilde{k}}\) and their complex conjugate are
proportional to
\begin{equation}\label{diagcondition}
\mathscr{D}(\eta,\tilde{k})=\frac{\nu}{2}\phi_{-\tilde{k}}'\phi_{\tilde{k}}'+
\frac{\nu}{2}\left(\mu-9\left(\frac{a'}{a}\right)^{2}\right)
\phi_{-\tilde{k}}\phi_{\tilde{k}}.
\end{equation}
The condition \(\mathscr{D}(\eta,\tilde{k})=0\) can be
rewritten as
\begin{equation}\label{diag2}
\frac{\left(\phi'_{\tilde{k}}(\eta)\right)^{2}}{\left(\phi_{\tilde{k}}(\eta)\right)^{2}}
=9\left(\frac{a'(\eta)}{a(\eta)}\right)^{2}-\mu(\eta,\tilde{k}).
\end{equation}
It is clear that at any finite time $\eta>\eta_c$, this equation can be
fulfilled,  i.e., the Hamiltonian can be made diagonal in the Fock basis,
namely by a suitable choice of mode function $\phi_{\tilde{k}}$. For the
case of the creation time $\eta_c$ itself, however, let us recall from
Sec.\ref{poseinitial} that all mode functions \(\phi_{\tilde{k}}\) are
proportional to their derivative with a fixed proportionality constant,
see (\ref{functioninetaatcreation}) and (\ref{derivinetaatcreation}).
Therefore, (\ref{diag2}) cannot be fulfilled at the creation time
\(\eta_{c}\). This means that whatever mechanism determines the initial
behaviour of modes cannot be described in terms of a Fock basis and a
corresponding particle picture.

\subsection{Initial field fluctuations}\label{danielssoncriterion}
It has been proposed, see \cite{danielsson02}, that Planck scale physics
might imply that when modes are created they appear in a state that
minimizes the product of the fluctuations of the mode's field
 \(\hat{\phi}_{\tilde{k}}(\eta)\) and its conjugate momentum field
\(\hat{\pi}_{\tilde{k}}(\eta)\) so that the product of the uncertainties
reads:
%In \cite{danielsson02} the usual mode function (without modifications
%caused by a cutoff) has been studied and the consequences of the minimum
%uncertainty criterion on a particular choice of solutions were examined.
%It was shown that the minimum uncertainty vacuum agrees with the adiabatic
%vacuum to zeroth order only.
\begin{equation}\label{uncertainty}
\Delta\hat{\phi}_{\tilde{k}}(\eta_c)\cdot
\Delta\hat{\pi}_{\tilde{k}}(\eta_c)=\frac{1}{2}
\end{equation}
In our case here, we find that at all times $\eta>\eta_c$:
\begin{eqnarray*}
\Delta\hat{\phi}_{\tilde{k}}(\eta)&=&\left|\phi_{\tilde{k}}(\eta)\right|\\
\Delta\hat{\pi}_{\tilde{k}}(\eta)&=&\nu(\eta,\tilde{k})~\left|\phi'_{\tilde{k}}(\eta)-3\frac{a'(\eta)}{a(\eta)}\phi_{\tilde{k}}(\eta)\right|
\end{eqnarray*}
 %The later can be written out as
%\begin{eqnarray*}
%\left|\nu\phi'_{\tilde{k}}-3\nu\frac{a'}{a}\phi_{\tilde{k}}\right|^{2}&=&\nu^{2}\cdot\left|\phi'_{\tilde{k}}\right|^{2}+9\nu^{2}\left(\frac{a'}{a}\right)^{2}\left|\phi_{\tilde{k}}\right|^{2}\nonumber\\
%&&-3\nu\frac{a'}{a}\left(\phi_{\tilde{k}}\phi_{\tilde{k}}^{'*}+\phi_{\tilde{k}}^{*}\phi'_{\tilde{k}}\right).
%\end{eqnarray*}
Let us now recall, see Sec.\ref{poseinitial}, that while
\(\phi_{\tilde{k}}\) and its derivative assume finite values for
\(\eta\rightarrow\eta_{c}\), the term  \(\nu(\eta,\tilde{k})\) diverges
\(\propto\left(\textnormal{plog}(-\beta\tilde{k}^{2}/a^{2})+1\right)^{-1}\).
On one hand, this means that the criterion of \cite{danielsson02} for
specifying the vacuum state is here not applicable. On the other hand, it
is perhaps not surprising that, while the modes are created with field
fluctuations \(\Delta\hat{\phi}_{\tilde{k}}\) of finite size, their
momentum field fluctuations \(\Delta\hat{\pi}_{\tilde{k}}\) are divergent
at the moment of mode creation itself.
%$\eta_{c}\).% and the plog-function equals \(-1\) at creation time \(\eta_{c}\).%At any other time \(\eta_{0}\neq\eta_{c}\), the criterion (\ref{uncertainty}) will lead to condition on the constants \(C_{1}\), \(C_{2}\) and therefore to a particular choice of vacuum. Note that \(C_{1}\), \(C_{2}\) are also constrained by the Wronskian condition (\ref{wronskian}).

\section{Ground state energy}\label{vacuumenergy}
%Before we derived the exact solution to the mode equation (\ref{equationofmotion}), we wrote down the Hamiltonian expressed through creation and annihilation operators (\ref{quantizedHamiltonian}) and
Since the Hamiltonian of a mode is quadratic in the field $\pi_{\tilde{k}}$, it is plausible that the divergence of the momentum field's fluctuations
\(\Delta\hat{\pi}_{\tilde{k}}\)
at $\eta_c$ should  imply a divergence of the expectation value of the mode's Hamiltonian at $\eta_c$.
To this end, let us consider the total ground state energy at some time \(\eta\):
\begin{eqnarray}
E_{vac}(\eta)&=&\int_{\tilde{k}^{2}<\frac{a^{2}}{e\beta}} d^{3}\tilde{k}\quad \rho_{vac,\tilde{k}}(\eta)\nonumber\\
&=&\int_{\tilde{k}^{2}<\frac{a^{2}}{e\beta}} d^{3}\tilde{k}\,
\bigg[\frac{\nu}{2}\left|\phi_{\tilde{k}}'\right|^{2}\nonumber\\
&&+\frac{\nu}{2}\left(\mu-9\left(\frac{a'}{a}\right)^{2}\right)
\left|\phi_{\tilde{k}}\right|^{2}\bigg]\label{Evac}
\end{eqnarray}
Recall that the integration ranges over only those modes whose
wavelength at time \(\eta\) exceeds the cutoff length.
At a given mode's creation time its operators $a_{\tilde{k}}$ and $a^\dagger_{\tilde{k}}$, first enter the total Hamiltonian, which means that the mode then first contributes to both ground state and dynamical energy. In particular, in an expanding universe, the total Hamiltonian continually picks up new ground state energy as it picks up new modes.

After inserting \(\mu(\eta,\tilde{k})\) and \(\nu(\eta,\tilde{k})\) into
(\ref{vacuumterm}) and making use of the fact that we know the exact behaviour of the solutions \(\phi_{\tilde{k}}(\eta)\) to the mode equation, we now indeed find a divergence of each mode's contribution to the total ground state energy at the mode's creation time. Concretely, straightforward calculation shows that each mode's Hamiltonian contains three types of divergent terms:
\begin{eqnarray*}
&\propto&\ln\left(\textnormal{plog}(-\beta\tilde{k}^{2}/a^{2})+1\right)\\
&\propto&1/\left(\textnormal{plog}(-\beta\tilde{k}^{2}/a^{2})+1\right)\\
 &\propto&1/\left(\textnormal{plog}(-\beta\tilde{k}^{2}/a^{2})+1\right)^{3}
\end{eqnarray*}
We obtain a simplified representation of the energetic behaviour of each mode by changing variables from the pair \((\eta,\tilde{k})\) to the new pair \((\tau,\tilde{k})\), and taking into account that
%the Jacobian factor
%\(\frac{d\tau}{d\tilde{k}}=\frac{2\tau}{(1+\tau)\tilde{k}}\)
the integration measure in (\ref{Evac}) transforms as
\begin{equation*}
d^{3}\tilde{k}=4\pi\tilde{k}^{2}d\tilde{k}=2\pi\tilde{k}^{3}\frac{(1+\tau)}{\tau}d\tau.
\end{equation*}
In these variables, each mode's ground sate energy contains only one divergent term, which behaves as \(1/(\tau+1)^{2}\) for \(\tau\rightarrow-1\).
We can now consider the behaviour of the entity that is of physical significance, namely the integrated ground state energy of all modes.

Here, we notice that the accumulation of ground state energy through mode
creation is offset to some extent by the fact that each individual mode's
contribution  to the total ground state energy
$\rho_{vac,\tilde{k}}(\eta)$, i.e., roughly speaking its
$\hbar\omega_{\tilde{k}}/2$, continually diminishes as its wavelength
expands, i.e., as its $\omega_{\tilde{k}}$ decreases.

Whether or not there is a net generating of ground state energy through the expansion depends crucially on the scale factor function $a(\eta)$ of the background spacetime. In the case of the background spacetime that we have considered so far, namely de Sitter space, we know that this calculation can only yield a constant result, since in its mode equation the dependence on $\tilde{k}$ can be eliminated, see (\ref{eofmexplicit}).

It should be very interesting, however, to carry out the integration of
the total ground state energy, at least numerically, for the case of a
more general background spacetime, such as the case of power-law
inflation. As already mentioned, we have calculated the exact solutions to
the exact mode equation in the power-law case, see Appendix
\ref{powerlaw}. The balance of the continual creation of ground state
energy and its continual dilution through expansion is nontrivial in any
model of quantum field theory in the sub-Planckian regime, including those
of generalized dispersion relations. This is of interest because if the
vacuum energy generation is not fully offset by its dilution then this
would imply a vacuum instability and therefore a potential starting
mechanism for inflation.

\section{Summary and outlook}

We considered quantum field theory in the sub-\-Planck\-ian regime, by
which we mean the regime of length scales larger than but close to the
Planck (or other cutoff-) length. In the case of an expanding background
spacetime the independent degrees of freedom of the QFT, i.e., the field
oscillators given by the comoving modes, are of particular interest. This
is because as the comoving modes' proper wavelength increases, new
comoving modes must continually enter the QFT description of the
sub-Planckian regime. A key problem is to identify in which state the
comoving modes first enter the QFT description of the sub-Planckian regime
and how they evolve through the sub-Planckian regime into the regime where
ordinary QFT holds.

To this end, we here followed up on a concrete model for quantum field
theory in the sub-Planckian regime. In this model a type of natural UV
cutoff is implemented that has been motivated from general quantum gravity
arguments, namely the presence of a finite minimum uncertainty in
positions. Previous work had found that in this model the initial phase of
a mode's evolution is described by a mode equation with an intriguing but
difficult to handle singular point at each mode's starting time.

Here, we improved on the existing numerical and semi-analytical solutions
by calculating the set of exact solutions to the precise mode equation for
the cases of de Sitter space and power-law inflation. In both cases the
initial singularity yielded for the roots of the indicial equation the
values $(0,3)$. This showed that the qualitative behaviour that can be
read off from of our exact solutions around the initial singularity is not
dependent on the precise dynamics of the scale factor. This in turn meant
that we were able to focus most of our calculations on the simpler case of
de Sitter expansion.

Having found the exact solution space, we studied possibilities for
identifying the physical mode solution within that solution space and
therefore for identifying the initial vacuum. The vacuum determines
 the mode's late time behaviour at horizon crossing and
therefore the type and magnitude of potentially observable effects in the
CMB.

Clearly, within any model for QFT in the sub-Planckian regime, it is
nontrivial to find a reliable method for identifying the physical mode
solution. In particular, given any natural UV cutoff it is of course no
longer possible to identify a mode's vacuum as having started out
essentially as the Minkowski vacuum on the basis that the mode would have
had \it arbitrarily \rm short wavelength in the distant enough past.

Within the model of QFT in the sub-Planckian regime that we studied here,
we found that indeed also any approach that is based on instantaneous
Hamiltonian diagonalization in a Fock basis must run into difficulties.
The reason is that, as we were able to establish using the exact
solutions, each mode's Hamiltonian enters the total Hamiltonian such that
it is at the mode's creation time not diagonalizable by means of any Fock
representation.

Similarly, the properties of the exact solutions to the mode equation
showed that the criterion for identifying the physical mode function by
minimizing the field uncertainty relation at mode creation time
 cannot be applied here unchanged: we found that
the field modes themselves are created with finite fluctuations. The
canonically conjugate momentum field's fluctuations are, however,
divergent at creation time itself, which is plausible given that the
momentum field generates changes in the field.

By studying the behaviour of the exact solutions close to the modes'
creation time we traced the underlying mathematical reason for why it is
difficult to give physical criteria that could reliably single out the
physical mode solution. Namely, we found that even though the mode
equation is second order in time, at creation time a mode solution cannot
be specified as usual by giving its amplitude and derivative. This is
because, at the creation time, the amplitude and derivative of all
solutions are proportional with the same proportionality constant.

This shows that in any model for QFT in the sub-Planckian regime the
 physical criteria for specifying the mode function and vacuum may need to
be adapted to unexpected mathematical behaviour. Here, for example, it is
possible to specify every choice of mode solution by specifying its
amplitude at creation time, by using the Wronskian condition and by also
using the freedom of overall phase to set the amplitude real at some
arbitrary time other than the mode's creation time.

Of course it should be most interesting to try to calculate directly from
candidate quantum gravity theories, such as loop quantum gravity and
string theory, the mode solution, i.e., to find the state in which these
theories predict new comoving modes to enter the sub-Planckian regime in
which the framework of QFT is valid. It would be most satisfying then if
the fixing of the initial state of the new modes could then be rephrased
within QFT in terms of a suitable boundary action along the lines of
\cite{schalmetal0401,schalmetal0411}. Within the model of QFT in the
sub-Planckian regime that we studied here, boundary terms have been
considered in \cite{amjad} and those results are likely to be useful for
this purpose.

Independent of which initial condition is to be chosen, our exact
solutions can be used to straightforwardly calculate from any initial
condition the behaviour of the corresponding mode function at horizon
crossing. In other words, the explicit solution space provides an explicit
bridge between the nontrivial initial conditions that may be set by Planck
scale physics and the predictions for their potential impact in the CMB.

We close with a gedanken experiment that indicates that Planck scale
physics may not necessarily set initial conditions for modes that enter
the sub-Planckian regime during expansion, as this may violate unitarity:
consider a hypothetical background spacetime that repeatedly contracts and
expands, so that comoving modes repeatedly enter and leave the
sub-Planckian regime of validity of QFT. In a contracting phase, when a
comoving mode's wavelength drops below the cutoff length, the field
operators of that mode drop out of the QFT's Hamiltonian. This means that
the further evolution of that mode is frozen, at least as far as the QFT
is concerned. Whatever excitation or particles that mode might have had
are then unaccessible within the framework of QFT because the quantum
field simply no longer contains operators that act nontrivially on the
dimensions of that mode's Hilbert space.

Thus, for all practical purposes any matter in such a comoving mode would
be disappearing as if behind a Planck ``horizon". This would be the case
until in a subsequent expansion phase the comoving mode's wavelength again
exceeds the cutoff length. Then also the Hamiltonian resumes a nontrivial
action on that mode's part of the Hilbert space. As far as the framework
of QFT is concerned, if the mode froze while in an excited state it will
re-emerge in the same excited state during expansion: the evolution is
unitary and no information was lost\footnote{It should be worth
investigating if a similar Hilbert space mechanism to the one discussed
here may freeze degrees of freedom that fall into a black hole, to thaw
them in the final evaporation of the black hole, thus preserving unitarity
without having the frozen degrees of freedom unduly gravitate.}. It is
possible, of course, that the full theory of quantum gravity will show
that there are operators which act on those frozen modes' dimensions in
the Hilbert space so that in a re-expanding phase, when the modes re-enter
the description by the framework of QFT, they do so with certain fixed
initial conditions. Clearly, in this case unitarity would be hard to
maintain in a cycle of expansions and contractions. If, however, Planck
scale physics were not to enforce a fixed initial condition on modes, then
the question remains unanswered in which state comoving modes are created
when they first emerge, during the very first expansion.

This leads to the question whether in the solution space to the mode
equation there exists any distinguished or canonical solution that might
therefore be the preferred physical solution for modes that emerge for the
first time, i.e., the question is if the mathematics singles out a
preferred vacuum state. This amounts to asking if there exists a canonical
splitting of the solution space that could improve on the useful but at
best approximate splitting of the solution space into positive and
negative frequency solutions during periods of adiabatic evolution. Here
we notice that, equivalently, the question is whether there is a canonical
split of the solution space into what in an adiabatic phase would be the
sine and cosine basis solutions to the mode equation (from which one would
then obtain positive and negative frequency solutions straightforwardly).
Because of the peculiar singularity at the mode creation time this is
indeed possible: recall that there exists a distinguished dimension of the
solution space, consisting of the real-valued mode solutions that vanish
at creation time. This induces a canonical ON basis in the two-dimensional
solution space that could be viewed as the generalizations of the sine and
cosine functions. It will be difficult to orthonormalize these functions
in practice because the exact solution's power series is needed in an
integration over all time while the power series is slow to converge at
late times. Nevertheless, it will be very interesting to investigate how
close the so-defined positive frequency mode solution would be to the
adiabatic vacuum in the adiabatic phase.

\subsection*{Acknowledgements}
This work was partially supported by PREA, CFI, OIT and the Canada
Research Chairs program of the National Science and Engineering Research
Council of Canada. LL acknowledges support by the International Council
for Canadian Studies during the early phase of this work and current
support by the DAAD.

\bibliographystyle{h-physrev}
\bibliography{thesisreferences}

\begin{appendix}
\section{The Frobenius method}\label{frobenius}
Consider  a differential equation of the
form
\begin{equation}\label{frobeniusform}
x^{2}\cdot y''(x)+x\cdot\left[x\cdot p(x)\right]\cdot
y'(x)+\left[x^{2}\cdot q(x)\right]\cdot y(x)=0,
\end{equation}
where \(x=0\) is a regular singular point, i.e., the coefficients have
convergent Taylor series expansions
\begin{eqnarray*}
x\cdot p(x)&=&\sum_{n=0}^{\infty}
p_{n}x^{n}\\
\textnormal{and }x^{2}\cdot q(x)&=&\sum_{n=0}^{\infty} q_{n}x^{n}
\end{eqnarray*}
in some interval \(|x|<\rho\). The Frobenius method states that then there exists a so-called indicial equation:
\begin{equation}\label{characteristic}
r\cdot(r-1)+p_{0}\cdot r+q_{0}=0
\end{equation}
Let \(r_{1}\) and \(r_{2}\) be the two roots of this equation, with \(r_{1}\geq
r_{2}\), if \(r_{1}\) and \(r_{2}\) are real. Then (\ref{frobeniusform})
has either on the interval \(-\rho<x<0\) or on \(0<x<\rho\) one solution
of the form
\begin{equation}\label{sol1}
y_{1}(x)=|x|^{r_{1}}\sum_{n=0}^{\infty}a_{n}(r_{1})\cdot x^{n}.
\end{equation}
In the case where the two roots \(r_{1}\) and \(r_{2}\) of
(\ref{characteristic}) differ by an integer (so that \(r_{1}-r_{2}=N\)), the second solution takes
the form
\begin{equation}\label{sol2}
y_{2}(x)=A\cdot y_{1}(x)\cdot\ln|x|+
|x|^{r_{2}}\sum_{n=0}^{\infty}c_{n}(r_{2})\cdot x^{n}.
\end{equation}
The constant \(A\) as well as the \(a_{n}(r_{1})\), \(c_{n}(r_{2})\) can be
determined by inserting each of the solutions into the equation
(\ref{frobeniusform})\footnote{Once having found the coefficients
\(a_{n}\) of the first solution, \(A\) and the \(c_{n}\) can also be
calculated from them; see Sec.\ref{solutiondesitter} and Appendix \ref{frictionless}.}. The solutions
converge at least for \(|x|<\rho\) and define a function which is analytic
around \(x=0\).

\section{Exact solution to the frictionless mode equation}\label{frictionless}
As shown in \cite{kempfniemeyer01}, the field redefinition
\(\psi_{\tilde{k}}(\eta)=\sqrt{\nu}\phi_{\tilde{k}}(\eta)\) allows to eliminate the friction term in (\ref{equationofmotion}), so that we
arrive at an equation of the form
\(\psi_{\tilde{k}}''+\omega_{\tilde{k}}^{2}(\eta)\psi_{\tilde{k}}=0\) with
\begin{eqnarray*}
\omega_{\tilde{k}}^{2}(\eta)&=&\mu-6\left(\frac{a'}{a}\right)^{2}+\left(\frac{\nu'}{2\nu}\right)^{2}-3\left(\frac{a'\nu'}{a\nu}\right)\\
&&-3\left(\frac{a''}{a}\right)-\left(\frac{\nu''}{2\nu}\right).
\end{eqnarray*}
The Wronskian condition (\ref{wronskian}) also simplifies to
\begin{equation*}
\psi_{\tilde{k}}(\eta)\psi^{'*}_{\tilde{k}}(\eta)-\psi^{*}_{\tilde{k}}(\eta)\psi'_{\tilde{k}}(\eta)=i.
\end{equation*}
Here we determine the corresponding field redefinition for the case of the mode
equation formulated in the \(\tau\)-variable, namely equation (\ref{eofmexplicit}).

Making the ansatz
\(\tilde{\phi}_{\tilde{k}}(\tau)=F(\tau)\tilde{\psi}_{\tilde{k}}(\tau)\)
and inserting into equation (\ref{eofmexplicit}), we find
\(F(\tau)=\frac{\tau+1}{\tau^{5/4}}\). Recall that the range for the
variable \(\tau\) is \(-1\dots 0\). Therefore, for later convenience, we
take a factor of \(\sqrt{i}\) out of \(F(\tau)\) and write the field redefinition in \(\tau\) as
\begin{equation}\label{newtrafo}
\tilde{\phi}_{\tilde{k}}(\tau)=-\frac{\tau+1}{\sqrt{i}(-\tau)^{5/4}}\tilde{\psi}_{\tilde{k}}(\tau).
\end{equation}
After the transformation (\ref{newtrafo}), the equation of motion for the
new field \(\tilde{\psi}_{\tilde{k}}(\tau)\) is \(\tilde{\psi}_{\tilde{k}}''+\tilde{\omega}_{\tilde{k}}^{2}(\tau)
\tilde{\psi}_{\tilde{k}}=0\) with
\begin{eqnarray}
\tilde{\omega}_{\tilde{k}}^{2}(\tau)&=&\tilde{\mu}\eta'^{2}-3\left(\frac{\tilde{a}''}{\tilde{a}}\right)-6\left(\frac{\tilde{a}'}{\tilde{a}}\right)^{2}-3\frac{\tilde{a}'\tilde{\nu}'}{\tilde{a}\tilde{\nu}}\nonumber\\
&&+3\frac{\tilde{a}'}{\tilde{a}}\frac{\eta''}{\eta'}-\frac{5+\tau}{4\tau(1+\tau)}\cdot\left(\frac{\tilde{\nu}'}{\tilde{\nu}}-\frac{\eta''}{\eta'}\right)\nonumber\\
&&+\frac{5(\tau+9)}{16\tau^{2}(1+\tau)}\label{omega2}.
\end{eqnarray}
Substituting \(x:=\tau+1\), the mode equation we need to solve
reads (note that the creation time now corresponds to \(x_{c}=0\)):
\begin{eqnarray}
\tilde{\psi}''_{\tilde{k}}-\frac{1}{16\beta\,{H}^{2}x^{2}(x-1)^{2}}\cdot\bigg(36\,\beta\,{H}^{2}{x}^{4}&&\label{trafowithouteofm}\\
-12\,\beta\,{H}^{2}{x}^{3}-27
\,\beta\,{H}^{2}{x}^{2}-24\,\beta\,{H}^{2}x&&\nonumber\\
+32\,\beta\,{H}^{2}+4\,{x}^
{3}-4\,{x}^{2}\bigg)\,\tilde{\psi}_{k}&=&0\nonumber
\end{eqnarray}
(The prime now denotes \(\partial/\partial\tau\).) Since this equation is of the form (\ref{frobeniusform}) with
\(\tilde{p}(x)=0\) and the desired behaviour for \(x^{2}\cdot
\tilde{q}(x)\), we can use the Frobenius method to solve it.

\subsubsection*{Application of the Frobenius method}
%\subsubsection{First Solution}
Our ansatz is
\(\tilde{\psi}_{\tilde{k}}(x)=x^{s}\sum_{n=0}^{\infty}b_{n}(s)\cdot
x^{n}\), which, sorted by powers of \(x\), leads to
\begin{eqnarray}
&&\sum_{n=0}^{\infty}b_{n}\left[(n+s)(n+s-1)-2\right]x^{n}\nonumber\\
&+&\sum_{n=0}^{\infty}b_{n}\left[-2(n+s)(n+s-1)+\frac{3}{2}\right]x^{n+1}\nonumber\\
&+&\sum_{n=0}^{\infty}b_{n}\left[(n+s)(n+s-1)+\frac{27}{16}+\frac{1}{4H^{2}\beta}\right]x^{n+2}\nonumber\\
&+&\sum_{n=0}^{\infty}b_{n}\left[\frac{3}{4}-\frac{1}{4H^{2}\beta}\right]x^{n+3}\nonumber\\
&-&\sum_{n=0}^{\infty}b_{n}\,\frac{9}{4}\,x^{n+4}=0.
\end{eqnarray}
The indicial equation in this case, found from the coefficient of
\(x^{0}\), reads
\begin{equation*}
s(s-1)-2=0,
\end{equation*}
with the two roots \(s_{1}=2\) and \(s_{2}=-1\). Note that again we find an integer difference between the two indicial roots, \(s_{1}-s_{2}=N=3\). Therefore the first solution is of the form (\ref{sol1}) with \(s_{1}=2\), while the second solution is of the form (\ref{sol2}) with \(s_{2}=-1\). We set the arbitrary normalization constant \(b_{0}=1\).\\
We further find that
\begin{eqnarray*}
b_{1}&=&\frac{4s(s-1)-3}{2(s(s+1)-2)},\\
b_{2}&=&\frac{1}{(s+1)(s+2)-2}\\
&&\times\bigg(\frac{(4s(s-1)-3)(4s(s+1)-3)}{4(s(s+1)-2)}\\
&&\qquad-s(s-1)-\frac{27}{16}-\frac{1}{4H^{2}\beta}\bigg),\\
b_{3}&=&\frac{1}{(s+2)(s+3)-2}\\
&&\times\bigg[\left(2(2+s)(s+1)-\frac{3}{2}\right)b_{2}\\
&&\qquad-\left(s(s+1)+\frac{27}{16}+\frac{1}{4H^{2}\beta}\right)b_{1}\\
&&\qquad+\left(\frac{1}{4H^{2}\beta}-\frac{3}{4}\right)\bigg].
\end{eqnarray*}
For \(n\geq 4\), we can establish a recursion formula for each coefficient
\(b_{n}\) as a function of its four predecessors\footnote{Note that here
the regression depth is greater by one than in the case of the
solution to (\ref{eofmexplicit}).} and \(s\):
\begin{eqnarray*}
b_{n}(s)&=&\frac{1}{(n+s)(n+s-1)-2}\\
&&\times\bigg[\left(2(n+s-1)(n+s-2)-\frac{3}{2}\right)b_{n-1}\nonumber\\
&&\qquad-\bigg((n+s-2)(n+s-3)\\
&&\qquad\qquad+\frac{27}{16}+\frac{1}{4H^{2}\beta}\bigg)b_{n-2}\nonumber\\
&&\qquad+\left(\frac{1}{4H^{2}\beta}-\frac{3}{4}\right)b_{n-3}\nonumber\\
&&\qquad+\frac{9}{4}\,b_{n-4}\bigg]\label{recursionwithout}
\end{eqnarray*}
To determine the \(b_{n}\) for the first solution, simply insert \(s=s_{1}=2\) in the expressions above.%; for the first three coefficients, this means:
%\begin{eqnarray}
%b_{1}&=&\frac{5}{8}\cdot b_{0}\\
%b_{2}&=&1/10\, \left( -1/4\,{\frac {1}{{H}^{2}\beta}}+{\frac {23}{8}} \right)
%\cdot b_{0}\\
%b_{3}&=&{\frac {1}{11520}}\,{\frac { \left( 585\,{H}^{2}\beta-300 \right)}{{H}^{2}\beta}}\cdot b_{0}
%\end{eqnarray}
%The first solution to the equation (\ref{trafowithouteofm}) then reads
%\begin{equation}
%\tilde{\psi}_{\tilde{k},1}(x)=
%x^{2}\sum_{n=0}^{\infty}b_{n}(s=2)x^{n}\label{wfirst}.
%\end{equation}
%\subsubsection{Second Solution}

The second solution is of the form
\begin{equation*}
\tilde{\psi}_{\tilde{k},2}(x)=B\cdot\tilde{\psi}_{\tilde{k},1}(x)\cdot\ln x+
x^{s_{2}}\sum_{n=0}^{\infty}d_{n}(s_{2})\cdot x^{n}
\end{equation*}
with \(s_{2}=-1\). %Again, as shown above for the case with friction, we have two ways of calculating the constant and the coefficients \(B\) and \(d_{n}\). First,
Using our results for the first solution \(\tilde{\psi}_{\tilde{k},1}\),
we find (\(N=3\))
\begin{equation*}
B=\lim_{s\rightarrow -1}(s+1)\cdot b_{N}(s)=\frac{1}{8H^2\beta}
\end{equation*}
and
\begin{equation*}
d_{n}(s=-1)=\left[\frac{d}{ds}\left(\left(s+1\right)b_{n}(s)\right)\right]_{s=s_{2}=-1}.
\end{equation*}
%From here we find:
%\begin{eqnarray}
%d_{1}&=&-\frac{5}{4}a_{0}\\
%d_{2}&=&\frac{1}{2}\left(\frac{29}{16}+\frac{1}{4H^{2}\beta}\right)a_{0}\\
%d_{3}&=&\left(-\frac{33}{16}-\frac{7}{24H^{2}\beta}\right)d_{0}
%\end{eqnarray}
%(The other method is to calculate \(B\) and the \(d_{n}\) directly by insertion.) So the second solution to (\ref{trafowithouteofm}) is
%\begin{equation}\label{wsecond}
%\tilde{\psi}_{\tilde{k},2}(x)=\frac{1}{8H^{2}\beta}\cdot x^{2}\sum_{n=0}^{\infty}b_{n}(s=2)x^{n} \cdot\ln x+x^{-1} \sum_{n=0}^{\infty}d_{n}(s=-1)x^{n}.
%\end{equation}
%\subsubsection{General Solution}
Again, the general solution to (\ref{trafowithouteofm}) is given by a
linear combination,
\begin{equation}\label{generalwithout}
\tilde{\psi}_{\tilde{k}}(x)=D_{1}\tilde{\psi}_{\tilde{k},1}(x)+D_{2}\tilde{\psi}_{\tilde{k},2}(x),
\end{equation}
where \(D_{1}\) and \(D_{2}\) are complex constants.%Again, we will examine these and their relation to \(C_{1},C_{2}\) later in this paper.

\subsubsection*{Wronskian condition}
To find the corresponding Wronskian condition, we plug the field redefinition into (\ref{newwronskian}). This leads to
\begin{eqnarray}\label{newtrafowronskian}
\tilde{\nu}(\tau,\tilde{k})\frac{1}{\eta'}\left(\frac{\tau+1}{(-\tau)^{5/4}}\right)^{2}&&\nonumber\\
\times\left[\tilde{\psi}_{\tilde{k}}(\tau)\tilde{\psi}'{}^{*}_{\tilde{k}}(\tau)-\tilde{\psi}^{*}_{\tilde{k}}(\tau)\tilde{\psi}_{\tilde{k}}'(\tau)\right]&=&i.
\end{eqnarray}
Splitting the constants \(D_{1}\) and \(D_{2}\) into their real and imaginary parts,
\begin{eqnarray*}
D_{1}&=&S_{1}+iL_{1}\qquad\textnormal{and}\\
D_{2}&=&S_{2}+iL_{2},
\end{eqnarray*}
imposing (\ref{newtrafowronskian}) at the creation time \(x_{c}=0\) leads to the condition:
\begin{equation}
S_{1}L_{2}-L_{1}S_{2}=\left(\frac{12H}{\beta^{3/2}\tilde{k}^{3}}\right)^{-1}
\end{equation}

%\subsubsection*{Equivalence of the solutions with and without friction}
%If we write the field redefinition in the form
%\(\sqrt{i}(1-x)^{5/4}\tilde{\phi}_{\tilde{k}}=-x\tilde{\psi}_{\tilde{k}}\),
%Taylor-expand \((1-x)^{5/4}\) around \(x=0\) and insert the two general
%solutions \(\tilde{\phi}_{\tilde{k}}\) and \(\tilde{\psi}_{\tilde{k}}\),
%we can show by comparison of the coefficients in front of equal orders of
%\(x\) that the constants are related by
%\begin{equation}\label{equivalence}
%D_{2}=-\sqrt{i}C_{2}.
%\end{equation}

\section{Exact solution for the general power-law case}\label{powerlaw}
The scale factor in a general power-law background reads
\begin{equation}\label{ageneral}
a(\eta)=\left(\frac{\eta}{\eta_{0}}\right)^{q},
\end{equation}
so that the inverse transformation \(\eta\rightarrow\tau\) (compare (\ref{deftau})) is given by
\begin{equation}
\eta=\eta_{0}\left(-\frac{\beta\tilde{k}^{2}}{\tau e^{\tau}}\right)^{\frac{1}{2q}}.
\end{equation}
%While the coefficients \(p_{p-l}(\eta,\tilde{k})\) and \(q_{p-l}(\eta,\tilde{k})\) for the equation of motion in \(\eta\) were given in (\ref{pp-l}) and (\ref{qp-l}), to find the corresponding quantities for the equation of motion in \(\tau\), we have to start from their formulations in (\ref{ptransformed}) and (\ref{qtransformed}) and insert the the power-law derivatives (\ref{derivp-l}) and (\ref{deriv2p-l}). \(\tilde{p}(\tau)\) and \(\tilde{q}(\tau)\) will now also dependent on the choice of  \(\eta_{0}\) the exponent \(q\):
%\begin{eqnarray*}
%\tilde{p}_{p-l}(q,\tau)&=&{\frac {6\,q+3\,q\tau+q{\tau}^{2}+1+2\,\tau+{\tau}^{2}}{ 2\left( 1
%+\tau \right) q\tau}}\\
%\tilde{q}_{p-l}(q,\eta_{0},\tau)&=&\frac{\frac{\tilde{k}^{2}\eta_{0}^{2}}{e^{\tau}} \left( -\frac {\beta\,\tilde{k}^{2}}{\tau\,
%e^{\tau}} \right) ^{\frac{1}{q}}-9\,{q}^{2}\tau+3\,{q
%}^{2}-6\,{q}^{2}{\tau}^{2}+3\,q+6\,q\tau+3\,q{\tau}^{2}}{4q^{2}\tau^{2}}
%\end{eqnarray*}
It can be shown that the exact mode equation (\ref{transformedeofm}) written for the power-law case still satisfies the necessary conditions for applying the Frobenius method. That is, the creation time \(x_{c}=0\) is a regular singular point of the explicit transformed mode equation
%also fulfill the conditions (\ref{conditionregular1}) and (\ref{conditionregular2}) for a regular singular point at \(\tau_{c}=-1\), therefore the Frobenius method can be used. We shift \(\tilde{p}_{p-l}(q,\tau)\) and \(\tilde{q}_{p-l}(q,\eta_{0},\tau)\) to \(x:=\tau+1\), insert them into the equation of motion (\ref{transformedeofm}) and arrive at
\begin{eqnarray}
x(x-1)^{2}\,\tilde{\phi}''_{\tilde{k}}&&\label{backgroundeofm}\\
+\left(\left(\frac{1}{2}+\frac {1}{2q}\right){x}^{3}-\frac {1}{2q}{x}^{2}+\frac{3}{2}x-2\right)\,\tilde{\phi}'_{\tilde{k}}&&\nonumber\\
+\,\tilde{G}(q,\eta_{0};x)\,\tilde{\phi}_{\tilde{k}}&=&0,\nonumber
 \end{eqnarray}
where we have used the abbreviation
\begin{eqnarray}
\tilde{G}(q,\eta_{0};x)&=&\left(\frac{3}{4q}-\frac{3}{2}\right)\,{x}^{3}
+\frac{3}{4}\,{x}^{2}\label{G}\\
&&+\bigg[\frac{3}{2}
+\frac{\tilde{k}^{2}\eta_{0}^{2}}{4{q}^{2} {e^{x-1}}}\nonumber\\
&&\qquad\times\left( -\frac {\beta\,\tilde{k}^{2}}{ \left( x-1
 \right) {e^{x-1}}} \right) ^{\frac{1}{q}}\bigg]\,x.\nonumber
\end{eqnarray}
We now work with the Frobenius ansatz
\begin{eqnarray}
\tilde{\phi}_{\tilde{k}}(q,\eta_{0};x)&=&\sum_{n=0}^{\infty}f_{n}(q,\eta_{0},u)\,x^{n+u}.\label{generalfrob}
%\tilde{\phi}'_{\tilde{k}}(q,\eta_{0};x)&=&\sum_{n=0}^{\infty}f_{n}(q,\eta_{0},u)(n+u)x^{n+u-1},\nonumber\\
%\tilde{\phi}''_{\tilde{k}}(q,\eta_{0};x)&=&\sum_{n=0}^{\infty}f_{n}(q,\eta_{0},u)(n+u)(n+u-1)x^{n+u-2},\nonumber
\end{eqnarray}
The \(f_{n}(q,\eta_{0},u)\) can be determined by inserting (\ref{generalfrob}) into (\ref{backgroundeofm}) and sorting by powers of \(x\). Since, however, in general the coefficient \(\tilde{G}(q,\eta_{0},x)\) need not be polynomial, we calculate its Taylor expansion around \(x_{c}=0\):
\begin{eqnarray*}
\tilde{G}(q,\eta_{0},x)&\approx&\tilde{G}(q,\eta_{0},0)\\
&&+\left( \frac{\tilde{k}^{2}{{\eta_{0}}}^{2}e}{4q^{2}} \left( \beta\,\tilde{k}^{2}e\right) ^{\frac{1}{q}}+{\frac {3}{2}} \right) x\\
&&+ \left( -\frac{\tilde{k}^{2}{{\eta_{0}}}^{2}e}{4q^{2}} \left( \beta\,\tilde{k}^{2}e\right) ^{\frac{1}{q}}+{\frac {3}{4}} \right) {x}^{2}\\
&&+\sum_{m=3}^{\infty}\frac{1}{m!}\left(\frac{\partial^{m} \tilde{G}(q,\eta_{0},x)}{\partial x^{m}}\right)_{x=0}x^{m}
\end{eqnarray*}
%Using this expansion, (\ref{backgroundeofm}) with the ansatz (\ref{generalfrob}) leads to the following results:
%\begin{eqnarray}
%&&\sum_{n=0}^{\infty}f_{n}(n+u)(n+u-3)x^{n-1}\nonumber\\
%&+&\sum_{n=0}^{\infty}f_{n}(n+u)\left(\frac{3}{2}-2(n+u-1)\right)x^{n}\nonumber\\
%&+&\sum_{n=0}^{\infty}f_{n}\left[(n+u)\left(n+u-1-\frac{1}{2q}\right)+\left(\frac{\partial \tilde{G}(q,\eta_{0},x)}{\partial x}\right)_{x=0}\right]x^{n+1}\nonumber\\
%&+&\sum_{n=0}^{\infty}f_{n}\left[(n+u)\left(\frac{1}{2}+\frac{1}{2q}\right)+\frac{1}{2}\left(\frac{\partial^{2} \tilde{G}(q,\eta_{0},x)}{\partial x^{2}}\right)_{x=0}\right]x^{n+2}\nonumber\\
%&+&\sum_{n=0}^{\infty}\sum_{m=3}^{\infty}f_{n}\frac{1}{m!}\left(\frac{\partial^{m} \tilde{G}(q,\eta_{0},x)}{\partial x^{m}}\right)_{x=0}x^{n+m}=0\label{backgroundinserted}
%\end{eqnarray}
%The first line of (\ref{backgroundinserted}) gives
Proceeding exactly as before (see Sec.\ref{eofmanalysis} and Appendix \ref{frobenius}), we find the same roots of the indicial equation as in the de Sitter case, namely \(u_{1}=3\), \(u_{2}=0\). Therefore, the structure of the solutions (i.e., one a mere polynomial, (\ref{sol1}), the other containing a logarithmic term, (\ref{sol2})) is the same in de Sitter case as well as in all power-law backgrounds.%; therefore the de Sitter model studied in the main part of our paper should already allow for conclusions for these more general cases.

The explicit expressions for the \(f_{n}(q,\eta_{0},u), n=1,2,3\) found from %the three following lines of (\ref{backgroundinserted}) read\footnote{We make use of (\ref{creationgeneralized}) to replace \(\eta_{0}^{2}\left(\beta\tilde{k}^{2}e\right)^{\frac{1}{q}}=\eta_{c}^{2}\).}:
inserting of the Frobenius ansatz are (where we set \(f_{0}=1\)):
\begin{eqnarray*}
f_{1}(u)&=& -\frac{u(7-4u)}{2(1+u)(u-2)}\\
f_{2}(q,\eta_{0},u)&=&-\frac{1}{(u+2)(u-1)}\bigg[u\left(u-1-\frac{1}{2q}\right)\\
&&\qquad-u\frac{(7-4u)(3/2-2u)}{2(u-2)}\\
&&\qquad+\frac{3}{2}+\frac{\tilde{k}^{2}{{\eta_{c}}}^{2}e}{4q^{2}}\bigg]\\
f_{3}(q,\eta_{0},u)&=&-\frac{1}{(3+u)u}\\
&&\times\bigg\{(2+u)\left(-2u-\frac{1}{2}\right)\cdot f_{2}(q,\eta_{0},u)\\
&&+\bigg[(u+1)\left(u-\frac{1}{2q}\right)\\
&&+\frac{3}{2}+\frac{\tilde{k}^{2}{{\eta_{c}}}^{2}e}{4q^{2}}\bigg]\cdot f_{1}(u)\\
&&+\left(\frac{3}{4}+u\left(\frac{1}{2}+\frac{1}{2q}\right)-\frac{\tilde{k}^{2}{{\eta_{c}}}^{2}e}{4q^{2}}\right) \bigg\}
\end{eqnarray*}
Here we used \(\eta_{c}\) as an abbreviation; in the power-law case \(\eta_{c}=\eta_{0}\left(\beta\tilde{k}^{2}e\right)^{\frac{1}{2q}}\).

%In the de Sitter case, we were able to determine a recursion formula for the \(n\)th coefficient based on a limited number of preceding coefficients. Contrary to this, i
In the general power-law background case the coefficient \(f_{n}(q,\eta_{0},u)\) for \(n\geq4\) depends on all the coefficients preceding it.
%before, that is, on \(f_{n-1}(q,\eta_{0},u),\,f_{n-2}(q,\eta_{0},u),\dots f_{1}(u),\,f_{0}\).
Explicitly, the recursion formula reads:
\begin{eqnarray*}
f_{n}(q,\eta_{0},u)&=&-\frac{1}{(n+u)(n+u-3)}\nonumber\\
&&\times\bigg\{(n+u-1)\nonumber\\
&&\qquad\times\left(\frac{3}{2}-2(n+u-2)\right)\cdot f_{n-1}\nonumber\\
&&+\bigg[(n+u-2)\left(n+u-3-\frac{1}{2q}\right)\nonumber\\
&&\qquad+\frac{3}{2}+\frac{\tilde{k}^{2}{{\eta_{c}}}^{2}e}{4q^{2}}\bigg]\cdot f_{n-2}\nonumber\\
&&+\bigg[(n+u-3)\left(\frac{1}{2}+\frac{1}{2q}\right)\nonumber\\
&&\qquad+\frac{3}{4}-\frac{\tilde{k}^{2}{{\eta_{c}}}^{2}e}{4q^{2}}\bigg]\cdot f_{n-3}\nonumber\\
&&+\sum_{m=4}^{m=n}\frac{1}{(m-1)!}\nonumber\\
&&\times\left(\frac{\partial^{m-1} \tilde{G}(q,\eta_{0},x)}{\partial x^{m-1}}\right)_{x=0}\cdot f_{n-m}\bigg\}
\end{eqnarray*}
%Whenever the Taylor series for \(\tilde{G}(q,\eta_{0},x)\) has a finite number of terms, this recursion formula will have with less than \(n\) regressions for the \(n\)th coefficient \(f_{n}\).%In general, however, the \(n\)th coefficient depends on all derivatives of  \(\tilde{G}(q,\eta_{0},x)\) up to the order of \(n-1\).
The first solution (\(u_{1}=3\)) in the general power-law background then reads
\begin{equation*}
\tilde{\phi}_{\tilde{k},1}(q,\eta_{0};x)=x^{3}\sum_{n=0}^{\infty}f_{n}(q,\eta_{0},u=3)\,x^{n}.
\end{equation*}
%where
%\begin{eqnarray*}
%f_{1}(u_{1}=3)&=&\frac{15}{8}\cdot f_{0},\\
%f_{2}(q,\eta_{0},u_{1}=3)&=&-\frac{1}{10}\left[3\left(2-\frac{1}{2q}\right)-32\frac{1}{4}+\frac{\tilde{k}^{2}{{\eta_{c}}}^{2}e}{4q^{2}}\right]\cdot f_{0},\\
%f_{3}(q,\eta_{0},u_{1}=3)&=&-\frac{1}{18}\bigg[-\frac{65}{2}\cdot f_{2}(q,\eta_{0},u_{1}=3)\\
%&&\qquad+\left(4\left(3-\frac{1}{2q}\right)+\frac{3}{2}+\frac{\tilde{k}^{2}{{\eta_{c}}}^{2}e}{4q^{2}}\right)\cdot f_{1}(u_{1}=3)\\
%&&\qquad+\left(\frac{3}{4}+\frac{3}{2}\left(1+\frac{1}{q}\right)-\frac{\tilde{k}^{2}{{\eta_{c}}}^{2}e}{4q^{2}}\right)\cdot f_{0}\bigg].
%\end{eqnarray*}
%and the other \(f_{n}\) are given by
%\begin{eqnarray*}
%f_{n}(q,\eta_{0},u_{1}=3)&=&-\frac{1}{(n+3)n}\bigg[(n+2)\left(\frac{3}{2}-2(n+1)\right)\cdot f_{n-1}\\
%&&\qquad+\left((n+1)\left(n-\frac{1}{2q}\right)+\frac{3}{2}+\frac{\tilde{k}^{2}{{\eta_{c}}}^{2}e}{4q^{2}}\right)\cdot f_{n-2}\\
%&&\qquad+\left(n\left(\frac{1}{2}+\frac{1}{2q}\right)+\frac{3}{4}-\frac{\tilde{k}^{2}{{\eta_{c}}}^{2}e}{4q^{2}}\right)\cdot f_{n-3}\\
%&&\qquad+\sum_{m=4}^{m=n}\frac{1}{(m-1)!}\left(\frac{\partial^{m-1} \tilde{G}(q,\eta_{0},x)}{\partial x^{m-1}}\right)_{x=0}\cdot f_{n-m}\bigg].
%\end{eqnarray*}
%Having calculated the first solution in a general power-law background, it is easy to find the second solution; it will be of the form
The second solution has the form
\begin{eqnarray*}
\tilde{\phi}_{\tilde{k},2}(q,\eta_{0};x)&=&F\cdot \tilde{\phi}_{\tilde{k},1}(q,\eta_{0};x)\cdot \ln x\\
&&+
\sum_{n=0}^{\infty}g_{n}(q,\eta_{0},u_{2}=0)\,x^{n},
\end{eqnarray*}
where the constant \(F\) can be calculated to read
\begin{eqnarray*}
F%&=&\lim_{u\rightarrow u_{0}} (u-u_{0})\cdot f_{N}(q,\eta_{0},u)\nonumber\\
%&=&\lim_{u\rightarrow 0} u\cdot f_{3}(q,\eta_{0},u)\nonumber\\
&=&\frac{\tilde{k}^{2}{{\eta_{c}}}^{2}e}{8q^{2}}.
\end{eqnarray*}
The coefficients \(g_{n}(q, \eta_{0},u=0)\) of the second power series are related to the \(f_{n}(q, \eta_{0}, u)\) by:
\begin{eqnarray*}
g_{n}(q,\eta_{0},u_{2}=0)%&=&\left[\frac{d}{du}\left(\left(u-u_{2}\right)\cdot f_{n}(q,\eta_{0},u)\right)\right]_{u=u_{2}}\nonumber\\
&=&\left[\frac{d}{du}\left(u\cdot f_{n}(q,\eta_{0},u)\right)\right]_{u=u_{2}=0}
\end{eqnarray*}
%Let us explicitly write down the first three coefficients:
%\begin{eqnarray*}
%g_{1}&=&0\\
%g_{2}&=&\frac{3}{4}+\frac{\tilde{k}^{2}\eta_{c}^{2}e}{8q^{2}}\\
%g_{3}&=&\frac{6+69q+3\frac{\tilde{k}^{2}\eta_{c}^{2}e}{q}}{144q}
%\end{eqnarray*}
%Again, we would like to stress that this repeats an important feature of the de Sitter case solution: even in a general power-law background, the coefficient \(g_{1}\) of the linear order in the second power series vanishes.
%So the second solution (\(u_{2}=0\)) is given by
%\begin{equation}
%\tilde{\phi}_{\tilde{k},2}(q,\eta_{0},x)=\frac{\tilde{k}^{2}{{\eta_{c}}}^{2}e}{8q^{2}}\cdot f_{0}\cdot{}x^{3}\cdot \ln x\cdot \sum_{n=0}^{\infty}f_{n}(q,\eta_{0},u_{1}=3)x^{n}+
%\sum_{n=0}^{\infty}g_{n}(q,\eta_{0},u_{2}=0)x^{n}.
%\end{equation}
The general solution for the power-law background case is given by the linear combination of the two solutions:
\begin{equation*}\label{powerlawgeneral}
\tilde{\phi}_{\tilde{k}}(q,\eta_{0};x)=E_{1}\,\tilde{\phi}_{\tilde{k},1}(q,\eta_{0};x)+E_{2}\, \tilde{\phi}_{\tilde{k},2}(q,\eta_{0};x)
\end{equation*}
\(E_{1},E_{2}\) are complex constants, constrained by the Wronskian condition, which in the power-law case becomes% They will have to obey the generalized Wronskian condition, which we will discuss in the next Section.
%\section{Generalized Wronskian condition }{\markboth{\hfill{}Chapter \thechapter. General power-law modified mode equation}{Section \thesection. Wronskian condition\hfill}}
%We noticed that the derivative \(\frac{\partial \tau}{\partial \eta}\) given in (\ref{derivp-l}) depends on the power-law parameters \(q, \eta_{0}\).
%Therefore now also the Wronskian condition as given by (\ref{newwronskian}) will pick up a dependence on \(q\) and \(\eta_{0}\); the Wronskian for the general power-law case is:
\begin{eqnarray}\label{generalizedwronskian}
-\frac{2q}{\eta_{0}}\,\frac{e^{\tau/2}\tau^{3}}{\beta^{2}\tilde{k}^{4}(1+\tau)^{2}}\,\left(-\frac{\tau e^{\tau}}{\beta\tilde{k}^{2}}\right)^{\frac{1}{2q}}&&\nonumber\\
\times\left[\tilde{\phi}_{\tilde{k}}(\tau)\tilde{\phi}
^{*}_{\tilde{k}}{}'(\tau)-
\tilde{\phi}^{*}_{\tilde{k}}(\tau)
\tilde{\phi}_{\tilde{k}}'(\tau)\right]&=&i.
\end{eqnarray}
%We will shift this expression to \(x:=\tau+1\), insert (\ref{powerlawgeneral}) and then derive a constraint on \(E_{1},E_{2}\) by looking at the creation time \(x_{c}=0\). To sort (\ref{generalizedwronskian}) by powers of \(x\) and find out the constant part remaining for \(x_{c}=0\), we will  have to use the series expansion of the prefactor\footnote{Note that in the de Sitter case (\(q=-1\), \(\eta_{0}=-1/H\)) the \(e\)-terms cancel out and the prefactor is only a polynomial.},
Using a Taylor expansion of the prefactor,
\begin{eqnarray*}
-\frac{2q}{\eta_{0}}\,\frac{e^{\tau/2}\tau^{3}}{\beta^{2}\tilde{k}^{4}(1+\tau)^{2}}\,
\left(-\frac{\tau e^{\tau}}{\beta\tilde{k}^{2}}\right)^{\frac{1}{2q}}\approx&&\\
\frac{2q\left(\frac{1}{e\beta\tilde{k}^{2}}\right)^{1/2q}}{\sqrt{e}\beta^{2}\tilde{k}^{4}
\eta_{0}}\cdot\frac{1}{x^{2}}
-\frac{5q\left(\frac{1}{e\beta\tilde{k}^{2}}\right)^{1/2q}}
{\sqrt{e}\beta^{2}\tilde{k}^{4}\eta_{0}}\cdot\frac{1}{x}+\dots,&&
\end{eqnarray*}
and splitting \(E_{1,2}\) into real and imaginary part,
\(E_{j}=T_{j}+iM_{j},\,j=1,2\), the condition we find from evaluating
(\ref{generalizedwronskian}) at the creation time \(x_{c}=0\) reads:
\begin{equation*}
T_{1}M_{2}-M_{1}T_{2}= \frac{\sqrt{e}\beta^{2}\tilde{k}^{4}\eta_{0}}{12q
\left(\frac{1}{e\beta\tilde{k}^{2}}\right)^{1/2q}}
\end{equation*}

\end{appendix}

\end{document}